\documentclass[]{interact}
\usepackage{epstopdf}

\usepackage[numbers,sort]{natbib}  
\bibpunct{[}{]}{,}{n}{,}{,}        
\renewcommand\bibfont{\fontsize{10}{12}\selectfont}

\usepackage[utf8]{inputenc}
\usepackage{textcomp}
\usepackage{graphicx}
\usepackage{amsmath}
\usepackage[version=4]{mhchem}
\usepackage{siunitx}
\usepackage{longtable,tabularx}
\title{Predicting the onset of airfoil stall in transitional and turbulent flows with a pressure-gradient-based separation model for RANS simulations}

\author{
\name{Kevin Patrick Griffin,$^1$\thanks{CONTACT K.~P. Griffin. Email: kevin.griffin@nlr.gov} Bumseok Lee,$^1$ Ganesh Vijayakumar,$^1$ Benjamin Savino,$^2$ 
Brett J. Bornhoft,$^{3,4}$ Omkar B. Shende,$^3$ and Michael P. Whitmore$^3$}
\affil{$^1$National Laboratory of the Rockies, Golden, CO, 80401.}
\affil{$^2$University of Mississippi, University, MS 38677.}
\affil{$^3$Stanford University, Stanford, CA, 94305.}
\affil{$^4$Cadence Design Systems Inc, San Jose, CA, 95134.}
}

\usepackage{subcaption}
\usepackage{floatrow}
\usepackage{todonotes}
\newcommand{\pder}[2]{\frac{\partial#1}{\partial#2}} 
\newcommand{\pderi}[2]{\partial#1 / \partial#2} 

\usepackage{glossaries}
\newacronym{dns}{DNS}{direct numerical simulation}
\newacronym{rans}{RANS}{Reynolds-averaged Navier-Stokes}
\newacronym{geko}{GeKO}{Generalized $k$-Omega}
\newacronym{les}{LES}{large-eddy simulation}
\newacronym{apg}{APG}{adverse pressure gradient}
\newacronym{fpg}{FPG}{favorable pressure gradient}
\newacronym{zpg}{ZPG}{zero pressure gradient}
\newacronym{bl}{BL}{boundary layer}
\newacronym{sst}{SST}{shear-stress transport}
\newacronym{nlr}{NLR}{National Laboratory of the Rockies}

\makeglossaries
\glsdisablehyper
\glsenableentrycount

\begin{document}

\maketitle

\begin{abstract}
We propose an improved version of the $k$-$\omega$ shear-stress transport Reynolds-averaged Navier-Stokes model that supersedes our prior work (Griffin et al. 2024 \emph{J. Turb.}) with an updated pressure-gradient sensor that generalizes well across geometries and to laminar, transitional, and turbulent flows alike. The model is based on a composite pressure gradient criterion that uses the solution of an intermittency transport equation to synthesize the separation criteria of the Falkner-Skan theory for laminar flows and the continuously critical theory for turbulent flows. The proposed model is applied to various airfoils including the thick airfoils representative of wind turbine applications (DU00-w-212, S809 and S814) and thinner airfoils used in aerospace applications [NACA23012 and NLF(1)-0416]. For all airfoils, the proposed model is found to preserve or improve the predictions of the baseline model for both fully turbulent and transitional airfoils as demonstrated with results for the lift and drag coefficients. The model is also found to improve the prediction of separation on the NREL Phase VI rotor in a preliminary study, suggesting that the model will generalize to other complex three-dimensional flows.
\end{abstract}

\textbf{Keywords-} 
k-omega SST turbulence model

\section{Introduction} \label{sec:intro}

The prediction of boundary-layer separation with \gls{rans} technology remains an outstanding research question with important aerodynamics applications to energy and transportation technologies. The $k$-$\omega$ \gls{sst} \gls{rans} model \citep{Menter1994,Menter2003} remains one of the most trusted and widely used open-source models for aerodynamics simulations in research and industrial applications despite well documented deficiencies in the prediction of flow separation \citep{Menter2003,Matyushenko2016,Zhong2018,Menter2019,Menter2021,griffin2025improved}. The prediction of separation is further complicated in transitional flows since the transition of a laminar boundary layer to turbulence can delay the onset of separation, the unsteady breakdown of a laminar separation bubble can lead to turbulent reattachment/transition and pressure gradients can greatly influence the onset of turbulence transition \citep{Crabtree1957}. 

In this study, an improved model for separation of transitional and turbulent boundary layers is proposed. The contributions of this paper are summarized as follows:
\begin{enumerate}
	\item Starting with the pressure-gradient sensor used in the transition model of Menter et al. \citep{Menter2015}, we extend the applicability of this sensor from laminar flows to transitional and turbulent flows by proposing a composite sensor using a linear mixing rule with the continuously critical theory for turbulent flows.
    \item Our composite pressure-gradient sensor is deployed as a separation correction to the $k$-$\omega$ \gls{sst} \gls{rans} model following the methodology proposed in our prior work \citep{griffin2025improved} but replacing the pressure non-dimensionalization of that work, which was based on global quantities, with our composite sensor. This has two advantages: (1) the present model does not require the specification of global velocity and length scales, which makes it more generalizable across geometries and easier to apply to complex geometries and (2) the present model is applicable to laminar, transitional, and turbulent flows.
    \item Our turbulence model is demonstrated to improve the prediction of separation on five airfoils and a three-dimensional wind turbine rotor including cases with forced and free transition.
\end{enumerate}

The remainder of this paper is structured as follows. In Section~\ref{sec:background}, the baseline \gls{sst} model and some relevant improvements are presented and discussed. In Section~\ref{sec:model}, the model formulation and computational setup are outlined. In Section~\ref{sec:results}, results for various airfoils and the NREL Phase VI rotor are presented. Finally, conclusions are drawn in Section~\ref{sec:conclusion}.

\section{Background} \label{sec:background}

For completeness, and to contextualize the contributions of the present work, we begin by presenting the state-of-the-art model for predicting separation and transition that will serve as our baseline model throughout this work. We consider the 1-eq $\gamma$ transition model of Menter et al. \citep{Menter2015}, which solves the transport equation for intermittency $\gamma$ 
\begin{equation} \label{eq:gamma}
	\pder{(\rho \gamma)}{t} + \pder{(\rho u_j \gamma)}{x_j} = P_\gamma - E_\gamma + \pder{}{x_j} \left[ \left( \mu + \frac{\mu_t}{\sigma_\gamma}  \right) \pder{\gamma}{x_j} \right],
\end{equation}
where $\rho$ is the density, $u_j$ are the components of the velocity, $\mu$ is the molecular viscosity, $\mu_t$ is the eddy (turbulent) viscosity and $\sigma_\gamma=1.0$. $P_\gamma$ and $E_\gamma$ are the production and destruction/relaminarization source terms of intermittency, respectively, as defined in Menter et al. \citep{Menter2015} and detailed in Appendix \ref{app:source_terms}. The intermittency solution enters the turbulence model by adjusting the kinetic energy production and destruction terms in the SST model and with an additional production term to make transition process more robust \citep{Menter2015}. The transport equations for the turbulent kinetic energy $k$ and the specific dissipation rate $\omega$ are
\begin{equation} \label{eq:k}
	\pder{\rho k}{t} + \pder{\rho u_j k}{x_j} = \widetilde{P} + P^{lim}-  \widetilde{D}+ \pder{}{x_j}\left[(\mu + \sigma_k \mu_t) \pder{k}{x_j}\right],
\end{equation}
\begin{equation} \label{eq:omega}
	\pder{\rho \omega}{t} + \pder{\rho u_j \omega}{x_j} = \frac{\alpha}{\nu_t}P - \beta \rho \omega^2 + \pder{}{x_j}\left[(\mu + \sigma_\omega \mu_t) \pder{\omega}{x_j}\right] + 2(1-F_1) \frac{\rho \sigma_{\omega 2}}{\omega} \pder{k}{x_j} \pder{\omega}{x_j},
\end{equation}
where $\widetilde{P}= \gamma P$, $\widetilde{D}= \max(\gamma, 0.1) D$,  the Kato-Launder source term $P=\mu_t S \Omega $ \cite{kato1993belaunder}, $D = \beta^* \rho \omega k$, $\beta^*=0.09$, $\nu_t=\mu_t/\rho$, $S$ is the strain rate magnitude, $\Omega$ is the  vorticity rate magnitude, and $P^{lim}$ is the additional production term. 
The remaining model coefficients are determined from a mixing rule based on the $F_1$ boundary-layer sensor as $\phi = F_1 \phi_1 + (1-F_1) \phi_2$,
where $\phi$ represents the parameters $\beta$, $\sigma_k$, $\sigma_\omega$ and $\alpha$, and the constants $\phi_1$ and $\phi_2$ are given as $\beta_1=0.075$, $\beta_2=0.0828$, $\alpha_1=5/9$, $\alpha_2=0.44$, $\sigma_{k1}=0.85$, $\sigma_{k2}=1.0$, $\sigma_{\omega 1}=0.49$ and $\sigma_{\omega 2}=0.856$.
$F_1$ is designed to tend to unity inside boundary layers and to zero outside and is defined as $F_1 = max(F_{1orig},F_3)$, where

\begin{equation} \label{eq:arg1}
    F_{1orig} = \tanh(arg_1^4)  
\end{equation}
\begin{equation} \label{eq:arg1}
	arg_1 = \min \left[ \max \left( \frac{\sqrt{k}}{\beta^* \omega d}, \frac{500 \nu}{d^2 \omega} \right), \frac{4\rho \sigma_{\omega 2} k}{CD_{k \omega} d^2} \right],
\end{equation}
\begin{equation} \label{eq:CD}
	CD_{k \omega} = \max \left( 2 \rho \sigma_{\omega 2} \frac{1}{\omega} \pder{k}{x_j} \pder{\omega}{x_j}, 10^{-10} \right),
\end{equation}
\begin{equation} \label{eq:arg1}
    F_3 = e^-(\frac{R_d}{120})^8,
\end{equation}
\begin{equation} \label{eq:arg1}
    R_d = \frac{\rho d \sqrt{k}}{\mu},
\end{equation}
$\nu=\mu/\rho$, and $d$ is the minimum distance to a wall.

The eddy viscosity is defined as
\begin{equation} \label{eq:mut}
	\mu_t = \frac{\rho a_1 k}{\max(a_1 \omega, S F_2)},
\end{equation}
where $a_1=0.31$, the strain rate magnitude $S = \sqrt{2 S_{ij} S_{ij}}$ and $F_2$ is a second boundary-layer sensor designed to tend to unity inside boundary layers and zero outside. $F_2 = \tanh(arg_2^2)$, where
\begin{equation} \label{eq:arg2}
	arg_2 = \max \left( 2\frac{\sqrt{k}}{\beta^* \omega d}, \frac{500 \nu}{d^2 \omega} \right).
\end{equation}

Eq.~(\ref{eq:mut}) was intended to switch from $\nu_t \sim k/\omega$ to $\nu_t \sim k/S$ in regions of strong \gls{apg}, assuming $a_1 \omega < SF_2$ is a good sensor for strong \gls{apg} regions \citep{Menter1993}. 
However, our prior work \citep{griffin2025improved} showed that the sensor is not well correlated with $\pderi{p}{s}$ (the streamwise derivative of pressure) and suggested direct use of $\pderi{p}{s}$ as a sensor for switching between two values of the $a_1$ coefficient. That sensor is nondimensionalized by the freestream velocity and a geometric length scale (e.g., the airfoil chord length). In complex external flows, such nondimensionalization of the pressure gradient by global constants will not accurately identify regions where the pressure gradient is strong or weak with respect to the local flow and turbulence, which may explain why the sensor did not generalize well (required recalibration) when applied to different geometries. In addition, the interaction of the sensor with a transition model has not been investigated. These issues are address in the present work.

\section{Model formulation and computational setup} \label{sec:model}
As in our prior work \citep{griffin2025improved}, the present model adds an explicit pressure gradient sensor to the $k$-$\omega$ SST model \citep{Menter2003}. Unlike our prior work, the strength of the pressure gradient is measured using the nondimensional parameter
\begin{equation}
	\lambda_\theta = -\frac{\theta^2}{\mu u_s}\pder{p}{s},
\end{equation}
where $s$ is the streamwise coordinate, $u_s$ is the streamwise component of velocity, and $\theta$ is the boundary-layer momentum thickness. In practice, this parameter is non-local and difficult to compute. Therefore, Menter et al. \citep{Menter2015} proposed a local approximation, calibrated with Falkner–Skan profiles at the boundary-layer midpoint, and applied clipping for robustness
\begin{equation}
	\lambda_\theta = \min \left[\max \left(-7.57\times 10^{-3} \pder{u_n}{n}\frac{d^2}{\nu}+0.0128, ~~ -1.0 \right), ~~ 1.0 \right],
\end{equation}
where $u_n$ is the wall-normal component of velocity and $n$ is the wall-normal coordinate, both of which can be readily computed by leveraging a baseline SST solver's infrastructure for computing $d$. The approximated parameter is used to detect flow separation in the laminar boundary layer in the present work. For laminar Falkner-Skan flows, flow separation occurs at a critical value of $\lambda_{\theta,l} = -0.0681$ \citep{Menter2015}. 

For turbulent flows with pressure gradients, separation has been observed to depend on the Reynolds number. Buri \citep{Buri1931} introduced the turbulent separation criterion $\lambda_{\theta,t} = \Gamma Re_\theta^{3/4}$, which is referred to as the continuously critical theory for turbulent flows. Subsequent studies confirmed this correlation with some revisions to the value of the constant $\Gamma$ \citep{Allan1963}. Consistent with the treatment of $Re_\theta$ in the Menter et al. \citep{Menter2015} transition model, $\theta \approx \delta/2$ and the model is assumed to be activated near the center of the boundary layer, which implies $\theta \approx d$, avoiding the additional implementation complexity of nonlocal operations to compute $\theta$ exactly. The velocity scale in $Re_\theta$ is the streamwise velocity magnitude, $u_s$, computed from the vector subtraction of the normal component of velocity from the local velocity, implying
\begin{equation}
	Re_\theta \approx \frac{\rho d u_s}{\mu}.
\end{equation}

To deploy the pressure gradient sensor in laminar and turbulent flows alike, we propose to use a linear mixing rule to arrive at the composite criterion
\begin{equation}
	\lambda_{\theta,c} = (1-\gamma)\lambda_{\theta,l} + \gamma \Gamma Re_\theta^{3/4},
\end{equation}
which satisfies the laminar and turbulent limits by using the intermittency $\gamma$ as determined by the solution of the transition model's transport equation in Eq.~(\ref{eq:gamma}). That is, the Falkner-Skan separation criterion is recovered in laminar flows and the continuous critical theory in turbulent flows. 
This proposed composite separation criterion is used to modify the $a_1$ coefficient in \gls{apg} regions following our prior work \citep{griffin2025improved}. Specifically, the baseline SST model's Eq.~(\ref{eq:mut}) is replaced with
\begin{equation}
	\mu_t =
	\begin{cases}
		\frac{\rho k}{\omega} & \text{if } a_1 \omega > S F_2, \\
		\frac{a_{1}' \rho k}{S F_2} & \text{otherwise},
	\end{cases}
	\label{eq:mut_mod}
\end{equation}
\begin{equation}
	a_{1}' =
	\begin{cases}
		a_{1,\mathrm{APG}} & \text{if } \lambda_{\theta} < \lambda_{\theta,c}, \\
		a_1 & \text{otherwise},
	\end{cases}
	\label{eq:a1_model}
\end{equation}
where $a_{1,\mathrm{APG}}$ is a reduced value of $a_1$, which promotes separation because lower eddy viscosity limits mixing of the high-momentum fluid at the boundary-layer edge with low-momentum fluid near the wall, which is susceptible to separation. Note that our prior work \citep{griffin2025improved} also explored independently modifying the $\beta^*$ coefficient but observed similar results as when $a_1$ was modified, so in this work, we choose to only modify $a_1$ since this term directly appears in the equation for eddy viscosity.

The model introduces two global coefficients that are iteratively calibrated with simulations of the S809 airfoil at Mach 0.1, a chord-based Reynolds number -- $Re_c$ -- of 2 million, and angles of attack -- $\alpha$ -- between 0 and 20 degrees. This airfoil has a 21\% thickness and is representative of the cross section of a wind turbine blade \citep{Somers1997_s809}. The values of $a_{1,\mathrm{APG}}=0.275$ and $\Gamma=-5\times 10^{-4}$ have been selected to achieve the best agreement with wind tunnel data. The accuracy and generalizability of this choice will be assessed in Section \ref{sec:results}.


\subsection{Computational setup for the airfoils}
\gls{rans} calculations are performed using ExaWind, an open-source high-fidelity software suite developed by the \gls{nlr} and Sandia National Laboratories that has been extensively verified and validated including verification on the convecting Taylor-Green vortex with convergence to the analytical solution established and validation through comparison to experimental data for various wind turbines and wind farms \citep{Sprague2020, Sharma2024}. ExaWind includes Nalu-Wind, a three-dimensional, incompressible, unstructured, unsteady, spatially second-order-accurate finite-volume solver. In this work, only two-dimensional calculations are performed. Solutions are time advanced for 10,000 time steps with a time step size of 2 steps per chord-pass time (chord length divided by freestream velocity) using an implicit second-order backward difference scheme \cite{Sharma2024}. This coarse time step is chosen since the focus of this work is on steady flow phenomena associated with the onset of separation rather than the unsteady large-scale separations that result at higher angles of attack. Even with this coarse time step, solution convergence with respect to time-step size is established in Appendix \ref{app:grid_conv} for all angles of attack. The solver is temporally advanced until the results are statistically stationary as measured by mean forces as described in Appendix \ref{app:force_conv}. Specifically, throughout the manuscript, time-averaged results are presented by averaging the samples from time steps 6000 through 10,000 since the lift and drag coefficients are essentially constant over this range. The linear solver used is the \textit{hypre} generalized minimal residual method with the BoomerAMG algebraic multigrid preconditioner \citep{hypre,HENSON2002155}. 

Airfoil calculations are performed using structured O-grids. 385 points each are distributed across the suction and pressure sides, with hyperbolic tangential stretching to achieve minimum spacings of $\Delta_t/c=6.1\times10^{-4}$ and $\Delta_t/c=3.3\times10^{-4}$ at the leading and trailing edges, respectively, where $c$ is the airfoil chord length. In addition, 162 points are geometrically distributed along the wall-normal direction to achieve a minimum spacing of $\Delta_y/c=2.1\times 10^{-7}$ with a stretching ratio of 1.13 and a domain radius of at least 600 chord lengths. For all cases, $\Delta_y^+<0.16$ as determined through a grid convergence study as discussed in Appendix \ref{app:grid_conv}. Additional details can be found in \cite{lee2025assessment,griffin2025improved}.

The transition simulations were conducted based on the experimental turbulence intensity of each case. First, the freestream values of $k$ and $\omega$ were calculated using the freestream velocity, the experimental turbulence intensity ($Tu$), and assuming $\mu_{t}/\mu=1$, which is an established best practice \cite{venkatachari2023implementation,hildebrand2023implementation,shankar2024verification}. Then, the experimental turbulence intensity was directly applied in the correlation formulations, $Re_{\theta c}=f(Tu,\lambda_{\theta})$, rather than using local values for $k$ and $\omega$.
Target turbulence intensities and Reynolds numbers for each airfoil considered are given in Table \ref{tab:airfoils}.

\begin{table}[]
	\begin{tabular}{llllll}
		\multicolumn{1}{l|}{\textbf{Airfoil name}}     &   \multicolumn{1}{l|}{Airfoil}  &  \multicolumn{1}{l|}{$Re_c$}   &  \multicolumn{1}{l|}{Tripped/smooth} & Turbulence  \\
        \multicolumn{1}{l|}{}     &   \multicolumn{1}{l|}{thickness}  &  \multicolumn{1}{l|}{}   &  \multicolumn{1}{l|}{} & intensity \\\cline{1-5} 
		\multicolumn{1}{l|}{\textbf{S809}~\citep{Somers1997_s809}} & \multicolumn{1}{l|}{21\%}           & \multicolumn{1}{l|}{$2.0\times 10^6$} & \multicolumn{1}{l|}{Both} &  0.04\%  \\
		\multicolumn{1}{l|}{\textbf{S814}~\citep{Somers1997_s814}} & \multicolumn{1}{l|}{24\%}           & \multicolumn{1}{l|}{$1.5\times 10^6$} & \multicolumn{1}{l|}{Both} & 0.04\% \\
		\multicolumn{1}{l|}{\textbf{DU00-w-212}~\citep{Pires2016}} &  \multicolumn{1}{l|}{21\%}            &  \multicolumn{1}{l|}{$3\times 10^6$}      &  \multicolumn{1}{l|}{Both} & 0.0864\% \\
		\multicolumn{1}{l|}{\textbf{NLF(1)-0416}~\citep{Somers1981}} & \multicolumn{1}{l|}{16\%}               & \multicolumn{1}{l|}{$4\times 10^6$}       & \multicolumn{1}{l|}{Both} & 0.15\%  \\
		\multicolumn{1}{l|}{\textbf{NACA23012}~\citep{Broeren2018}} & \multicolumn{1}{l|}{12\%}               &  \multicolumn{1}{l|}{$1.8\times 10^6$}       & \multicolumn{1}{l|}{Smooth only} & 0.1\%  \\
	\end{tabular}
	\caption{For each airfoil studied, the name, airfoil thickness as a percent of the chord length, chord-based Reynolds number, availability of tripped and/or smooth experimental data, and wind-tunnel turbulence intensity is listed.}
	\label{tab:airfoils}
\end{table}

\subsection{Computational setup for the NREL Phase VI rotor}
\begin{figure}
	\centering
	\floatsetup[figure]{style=plaintop} 
    \includegraphics[width=0.8\textwidth]{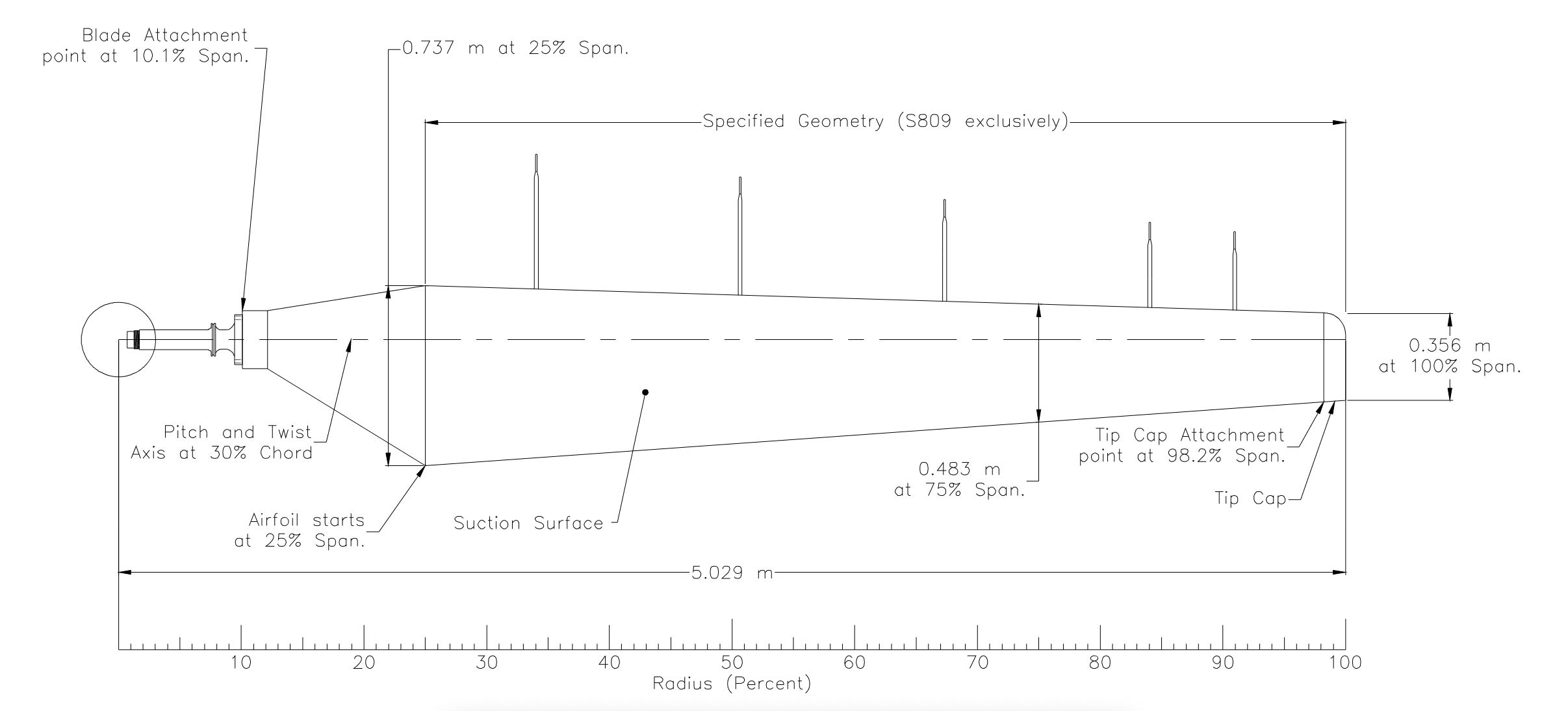}
	\caption{Planform of the NREL Phase VI wind turbine blade \cite{Hand2001}.}
	\label{fig:phase_vi_blade}
\end{figure}

The NREL Phase VI rotor is selected for validation since wind tunnel tests have been performed \citep{Hand2001}, the geometry is three dimensional (complementing the range of two-dimensional airfoils considered) and separation/transition phenomena are crucial to predict quantities of interest and are a challenge to predict with the baseline SST model. The NREL Phase VI rotor has two blades with a diameter of 10.06 meters. The blade section consists of a single airfoil, S809, starting from the 25\% radial location. A planform view of the rotor blade geometry is shown in Figure \ref{fig:phase_vi_blade} illustrating taper and sweep. 
The blade has a nonlinear twist distribution with a blade tip pitch of 3 degrees.

This simulation is performed using the ExaWind hybrid solver, which is composed of the unstructured Nalu-Wind solver for a body-fitted mesh on the rotor, the structured AMR-Wind solver for the background mesh, and overset grid assembler TIOGA for overset connectivity \cite{Sharma2024,kuhn2025amr}. The body-fitted mesh undergoes rigid-body rotation while the background mesh is stationary via overset meshing; fluid-structure interactions are not considered. 
The body-fitted (Nalu-Wind) mesh is an O-grid with 500 points on the airfoil surface, a near-wall resolution $\Delta_y^+ < 0.49$, and a stretching factor in the wall-normal direction of 1.15. The overset boundary is located 1.5 meters from the blade surface. The two blades are connected by a cylinder at the center, and other components, such as the spinner or tower, were not included in the CFD mesh. The total number of cells in the body-fitted mesh is 23,192,978.

The background (AMR-Wind) mesh was generated using the built-in capability of AMR-Wind. The domain ranges from -100 meters to 150 meters in the streamwise direction and -100 meters to 100 meters in the vertical and horizontal directions, with a grid size of 0.78125 meters. Four levels of static mesh refinement are applied near the rotor disk plane to capture vortical structures. The grid size at the refinement level four is 0.1953 meters and is distributed from -2.5 meters to 5 meters in the streamwise direction, -7 meters to 7 meters in the horizontal and vertical directions, resulting in a total of 45,527,040 cells.

\section{Results} \label{sec:results}

The proposed turbulence model is first applied to the two-dimensional airfoils summarized in Table \ref{tab:airfoils}. In the following results, the nomenclature in the figure labels is as follows: SST-$\gamma$ denotes the baseline model with the transition model presented in Section \ref{sec:background}; SST (without the -$\gamma$ suffix) denotes the fully turbulent version of the baseline model for which $\gamma=1$ and the $\gamma$ transport equation is not solved; the $\lambda_\theta$ suffix indicates that the proposed sensor in Eqs.~(\ref{eq:mut_mod}) and (\ref{eq:a1_model}) is added to either the SST or SST-$\gamma$ models. All experimental airfoil data is labeled as tripped or smooth depending on if the experimentalists forced transition with artificial roughness elements or if a nominally smooth surface is considered. In our simulations, we use the SST-$\gamma$ for smooth surfaces and SST for tripped surfaces, the latter of which assumes the trip leads to immediate transition starting at the leading edge. In practice, across the studies considered, various experimental tripping devices are employed at various streamwise locations; the effect of and sensitivity to these nuances is not considered in this work. 

\begin{figure}[h]
	\centering
	\floatsetup[figure]{style=plaintop} 
	\subcaptionbox{}{\includegraphics[width=0.49\textwidth]{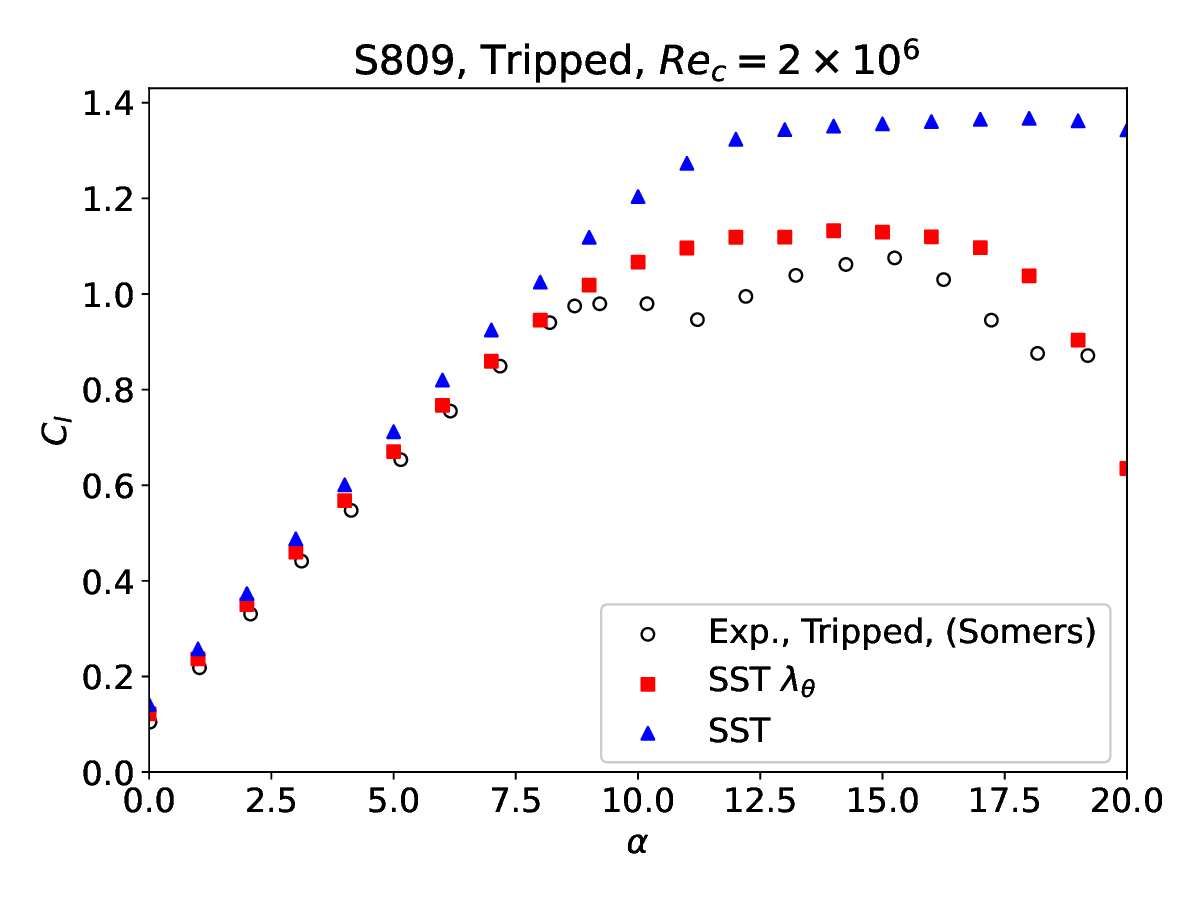}}
	\subcaptionbox{}{\includegraphics[width=0.49\textwidth]{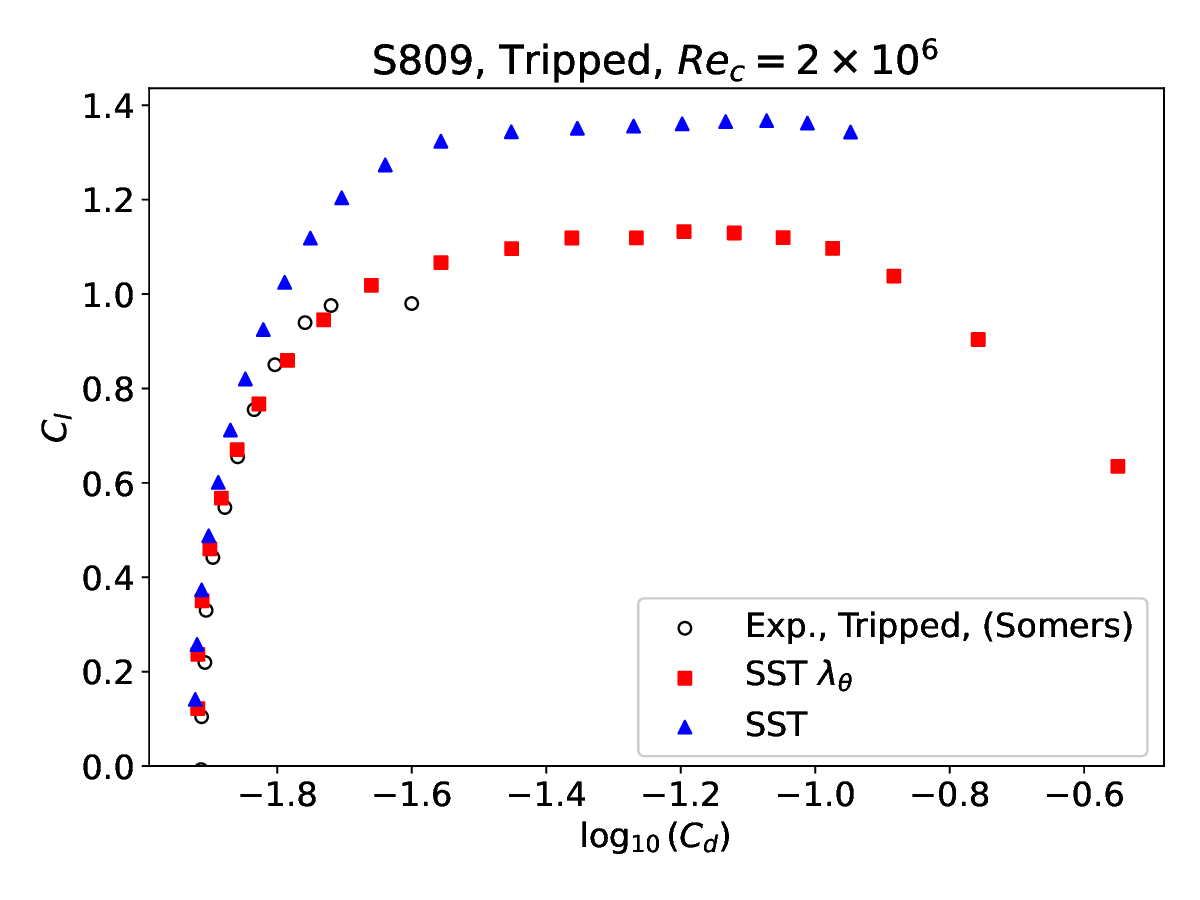}}
	\subcaptionbox{}{\includegraphics[width=0.49\textwidth]{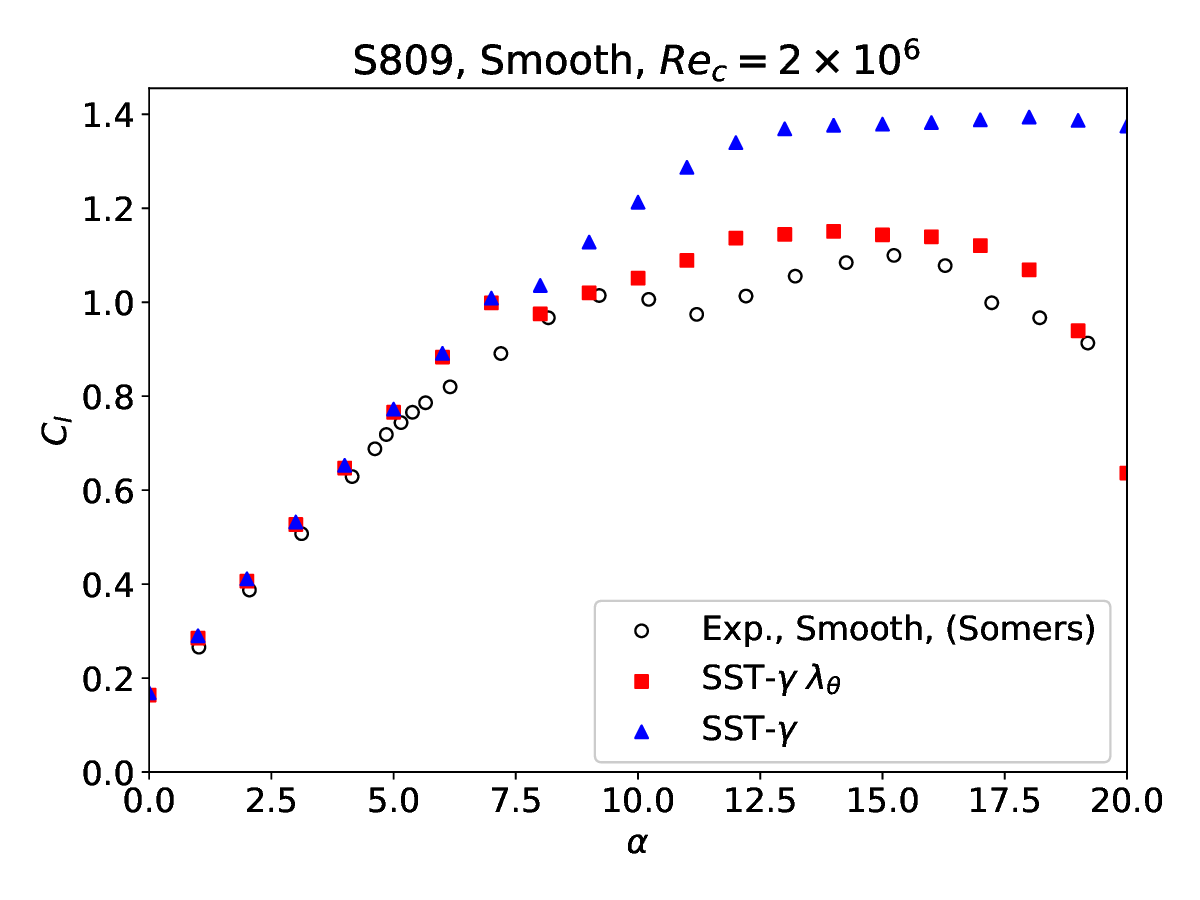}}
	\subcaptionbox{\label{fig:s809_cd_trans}}{\includegraphics[width=0.49\textwidth]{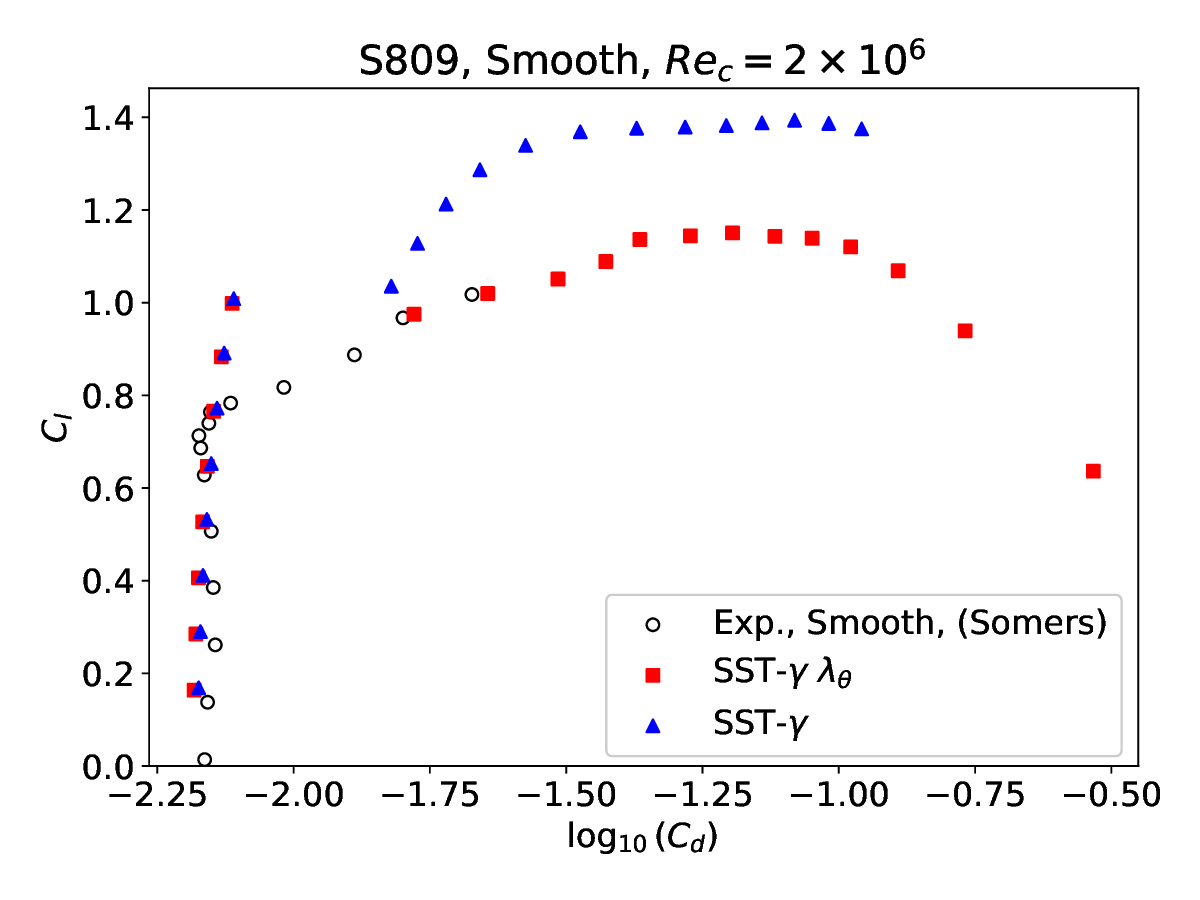}}
	\caption{Coefficient of lift versus (a,c) angle of attack and (b,d) coefficient of drag for the S809 airfoil for the (a,b) tripped and (c,d) smooth cases. The results for the proposed model (red) and baseline model (blue) are compared with experimental reference data (black) \citep{Somers1997_s809}.}
	\label{fig:s809}
\end{figure}

The first airfoil case considered is the S809 airfoil, which is a 21\% thick airfoil representative of a wind turbine blade cross section. Such airfoils are designed to exhibit a gradual aerodynamic stall, which manifests as a flattening of the slope of the lift versus angle of attack curve. As shown in Figure \ref{fig:s809}, the models are applied for both the tripped and smooth cases, which are modeled with and without the transition model, respectively. The simulation results that use the proposed $\lambda_\theta$ correction show lift predictions at high angles of attack that agree better with the experimental data than the predictions of the baseline model without the $\lambda_\theta$ correction. This indicates that our pressure-gradient-based sensor is improving the prediction of stall observed in the experimental data. For the prediction of drag, the data with the $\lambda_\theta$ correction is more accurate for both the tripped and smooth cases. 
However, for the smooth case at lift values between 0.8 and 1.0, the drag is underpredicted regardless of the separation correction, as shown in Figure~\ref{fig:s809}(d). Figure \ref{fig:s809}(c) indicates that these lift values correspond to angles of attack below the stall angle. The underprediction of drag below stall is attributed to a known limitation of the transition model. The model tends to predict a delayed reduction of the separation bubble, which leads to more laminar flows and larger laminar drag bucket~\cite{menter2022}.
For the smooth drag data corresponding to angles of attack greater than the stall angle, $\log_{10}(C_d)\ge-1.8$, there is a noticeable improvement in the drag prediction for the present $\lambda_\theta$ model compared to the baseline model. However, there is limited experimental data past stall for this case. For the tripped drag data, where uncertainties from the transition model are not present, the proposed correction improves the drag predictions at all angles of attack.

\begin{figure}[h]
	\centering
	\floatsetup[figure]{style=plaintop} 
	\subcaptionbox{Pressure coefficient}{\includegraphics[width=0.49\textwidth]{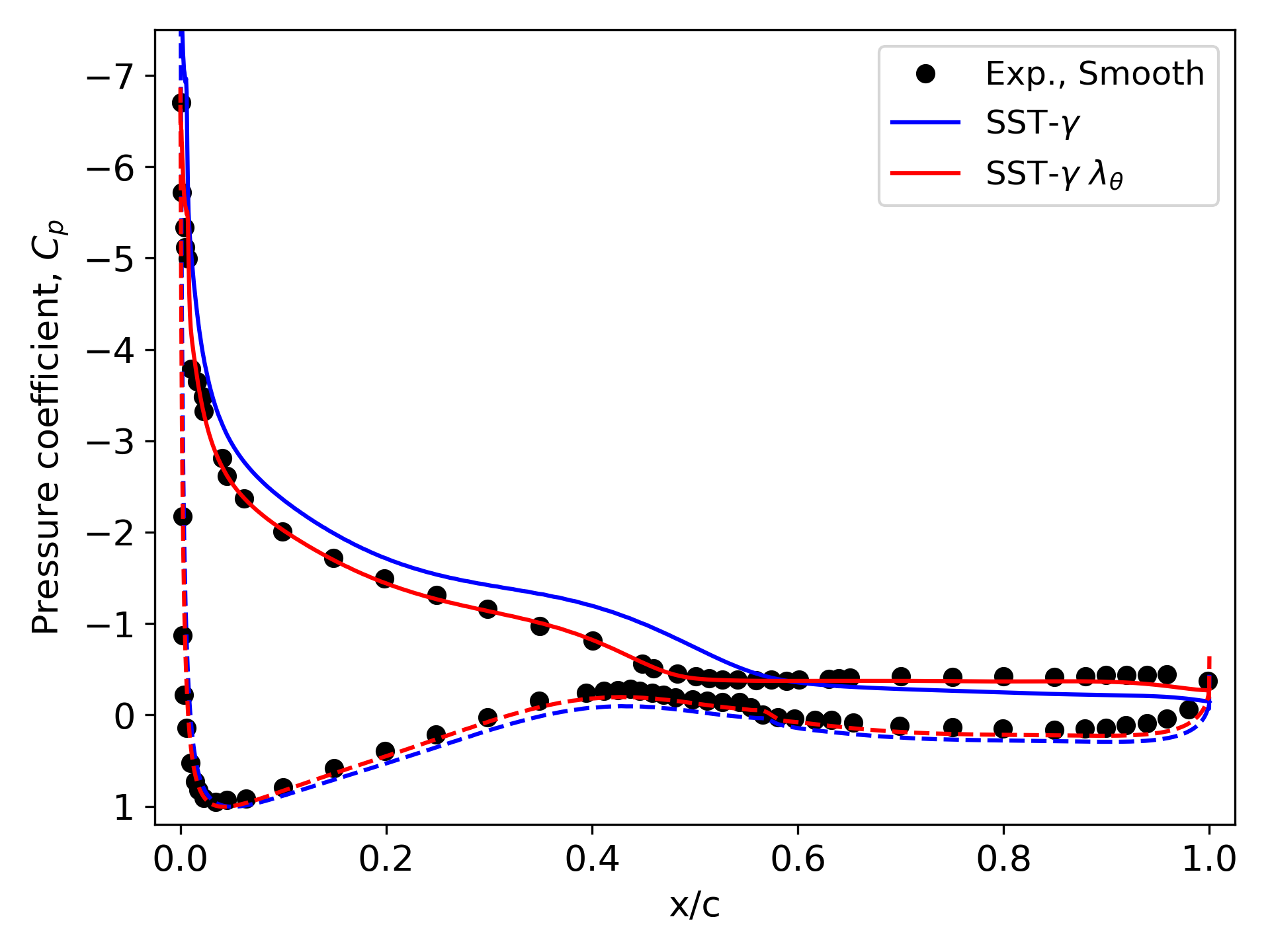}}
	\subcaptionbox{Skin friction coefficient}{\includegraphics[width=0.49\textwidth]{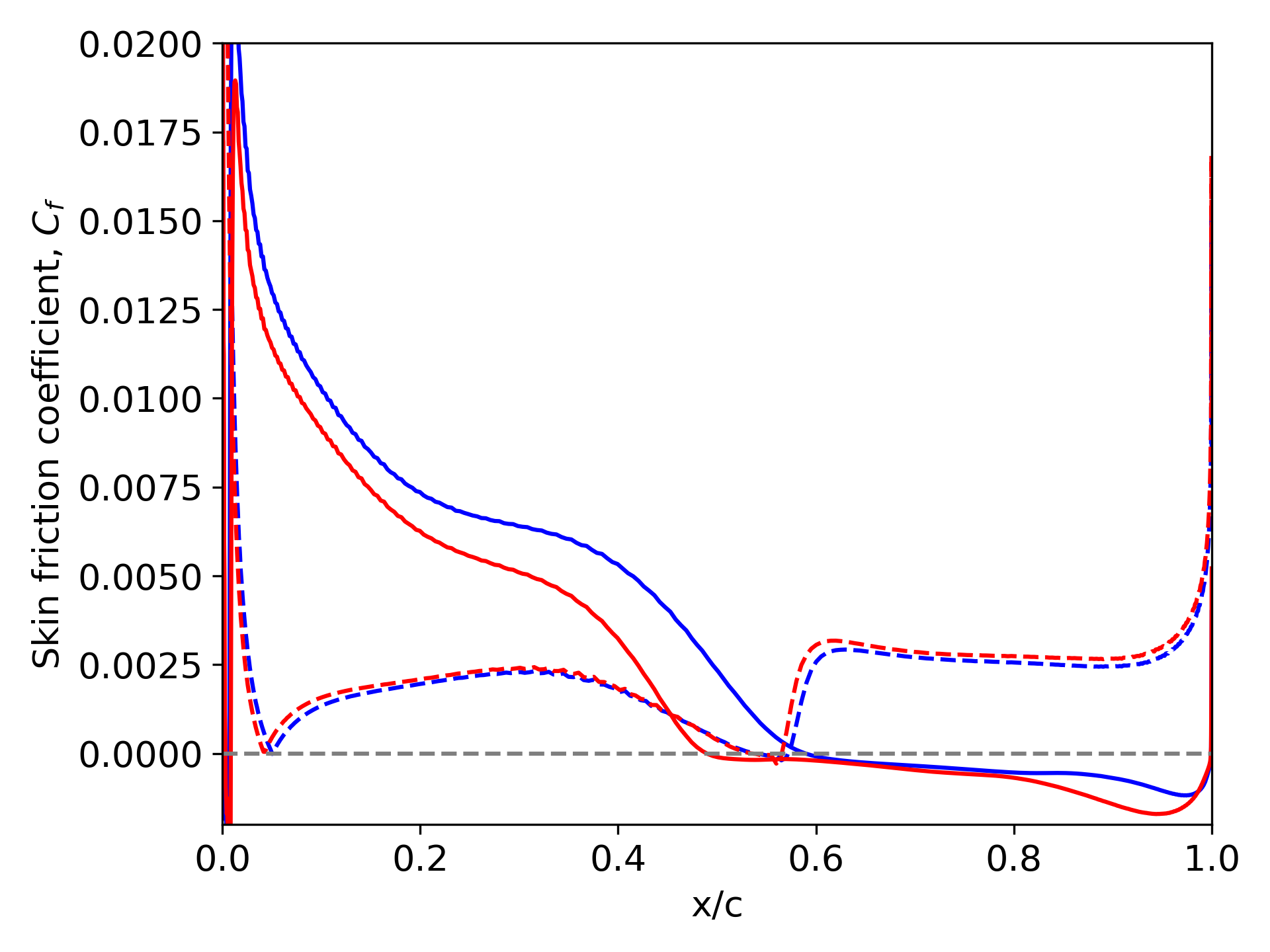}}
	\caption{Comparison of the (a) pressure and (b) skin friction coefficients at $\alpha=14^{\circ}$ for the smooth S809 airfoil. Solid/Dashed line: Upper/Lower surfaces, respectively.}
	\label{fig:s809_cpcf}
\end{figure}

\begin{figure}[h]
	\centering
	\floatsetup[figure]{style=plaintop} 
	\subcaptionbox{SST-$\gamma$}{\includegraphics[width=0.49\textwidth]{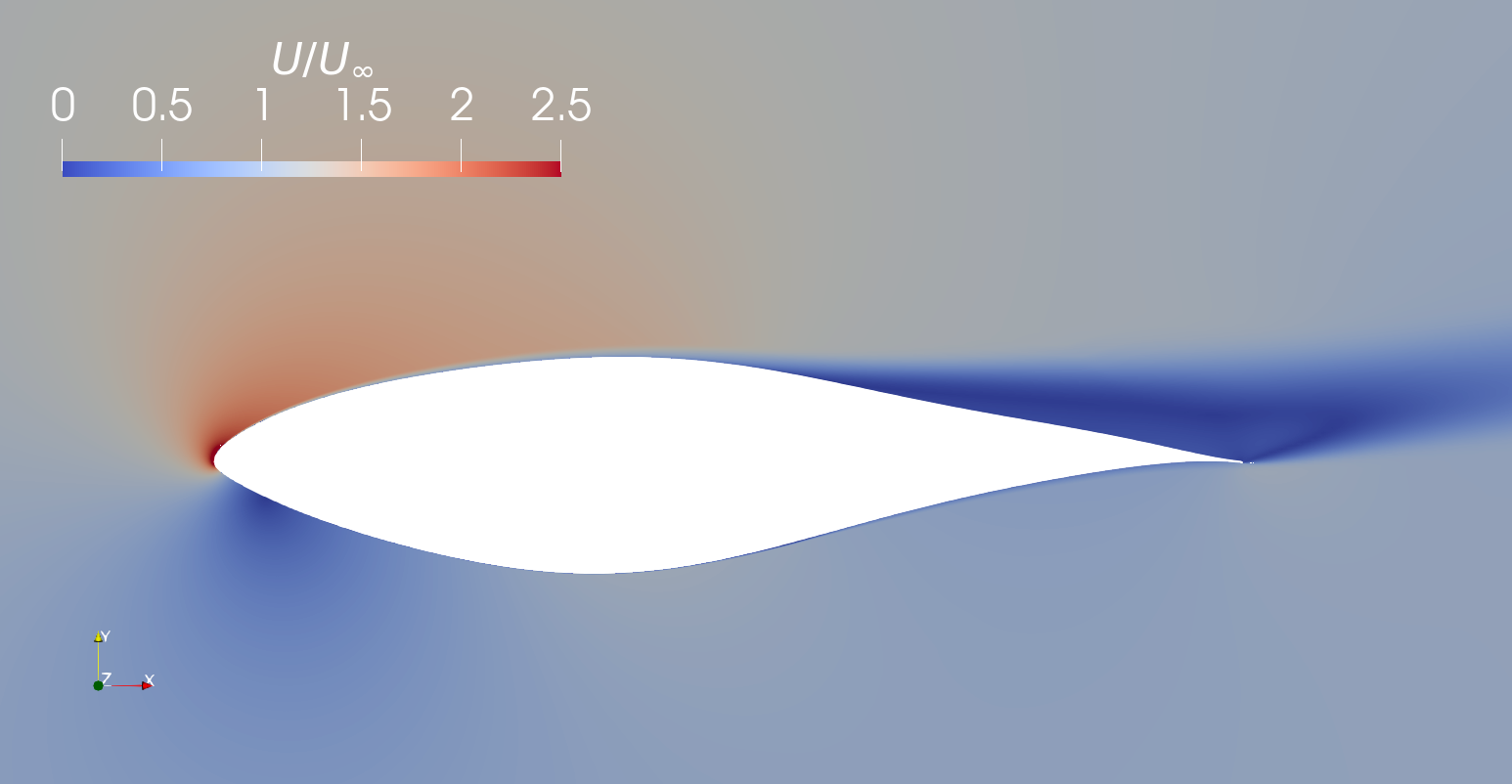}}
	\subcaptionbox{SST-$\gamma$ $\lambda_{\theta}$}{\includegraphics[width=0.49\textwidth]{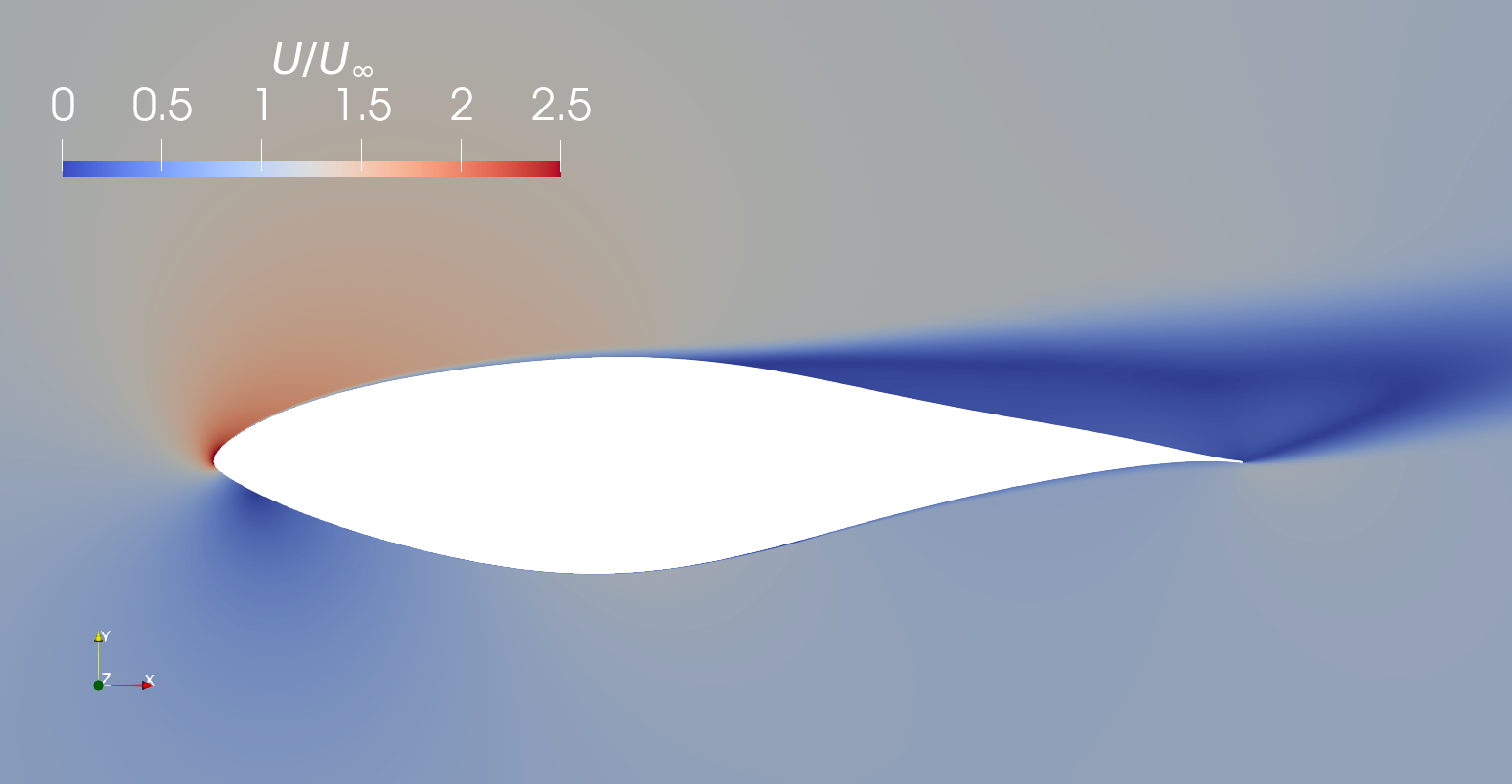}}
	\caption{Comparison of the velocity magnitude contours non-dimensionalized by the freestream velocity for the $\alpha=14^{\circ}$ for the smooth S809 airfoil.}
	\label{fig:s809_vel}
\end{figure}

The effects of the separation correction are further analyzed using the surface pressure distributions, skin friction coefficients, and velocity contours. Figure~\ref{fig:s809_cpcf} presents the surface pressure and skin-friction coefficients at $\alpha = 14^{\circ}$ for the smooth case. In the plots, the solid and dashed lines represent the upper and lower surfaces, respectively, while experimental measurements of the pressure coefficient are shown as black circles. The results with the correction show improved agreement with the experimental data over both surfaces, with particularly significant improvements on the upper surface. The skin friction coefficient indicates that flow separation occurs on the upper surface at $x/c = 0.49$ with the correction, compared to $x/c = 0.59$ without it. This result shows that the more accurate prediction of the separation location enhances agreement in the pressure distribution not only downstream of the separation region but also upstream. Experimental skin friction data is not available. At this high angle of attack, the influence of the laminar separation bubble is minimal, as transition occurs very close to the leading edge on the upper surface and only a small bubble forms on the lower surface. In other words, the laminar separation bubble has a more significant effect at lower angles of attack within the laminar drag bucket, where the separation correction has negligible impact. For detailed validation and analysis of the laminar separation bubbles at these lower angles of attack using surface pressure and skin friction data, the reader is referred to \cite{lee2025assessment}. Figure~\ref{fig:s809_vel} compares the non-dimensional velocity magnitude contours, normalized by the freestream velocity, for the same cases. The case with the correction shows a larger separated flow region on the upper surface, attributed to earlier flow separation with reduced turbulent eddy-viscosity production.

\begin{figure}[h]
	\centering
	\floatsetup[figure]{style=plaintop} 
    \includegraphics[width=0.6\textwidth]{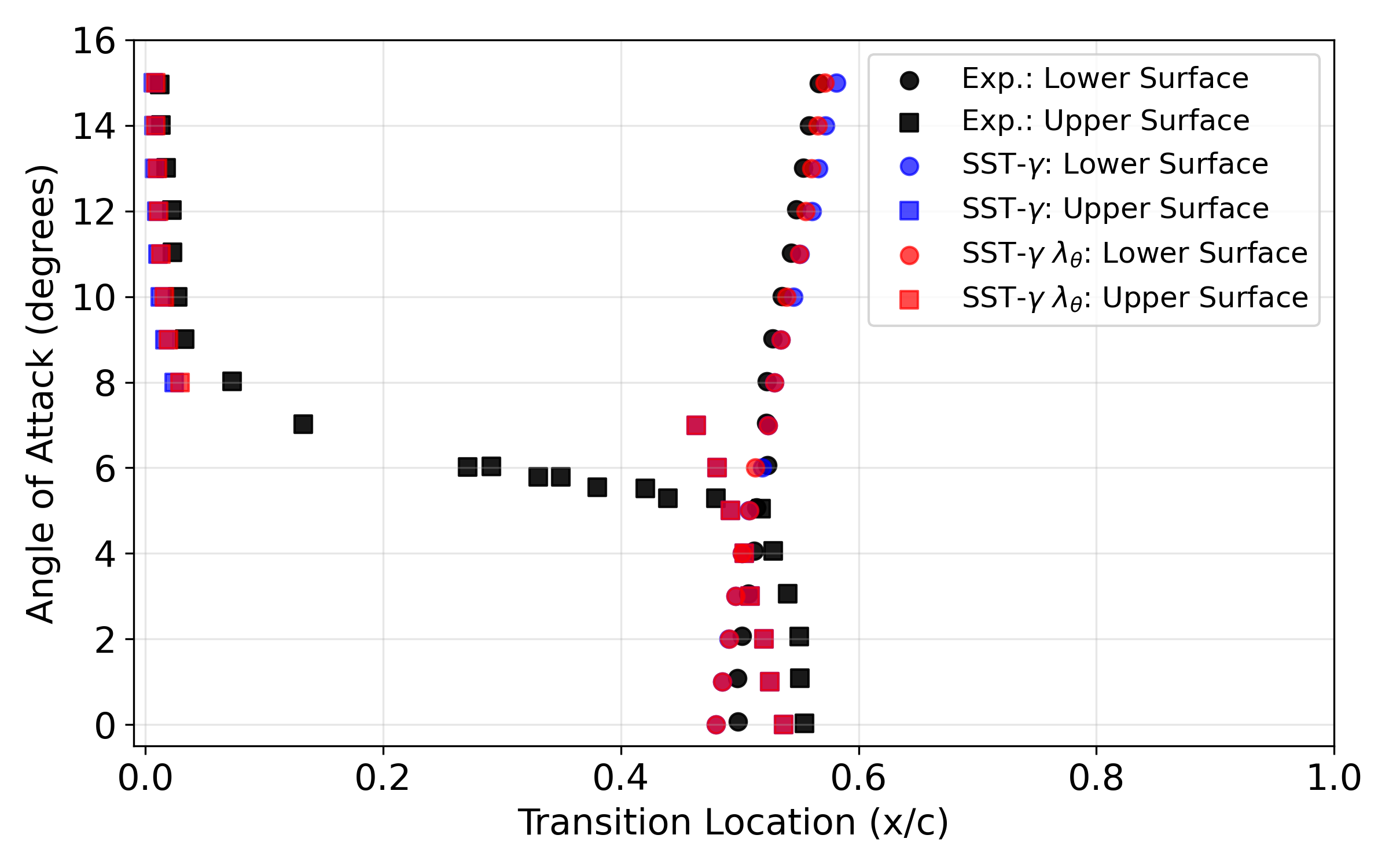}
	\caption{Comparison of the transition locations on the smooth S809 airfoil surface.}
	\label{fig:s809_trans_loc}
\end{figure}

For the S809 airfoil, the experimental data include transition locations, specifically the locations of turbulent reattachment. Figure~\ref{fig:s809_trans_loc} compares the transition locations for the smooth case among the SST–$\gamma$, SST–$\gamma$ $\lambda_{\theta}$ predictions, and the experimental measurements. In the figure, the transition locations on the lower and upper surfaces are represented by circle and square symbols, respectively. The results from the SST–$\gamma$ and SST–$\gamma$ $\lambda_{\theta}$ models show negligible differences up to an angle of attack of $7^{\circ}$ on both surfaces, as the correction has little influence in this range. At $\alpha=6^{\circ}$ and $7^{\circ}$, both models predict delayed transition compared to the experiment, leading to an underprediction of drag, as shown in Figure~\ref{fig:s809_cd_trans}. For $\alpha	\geq 8^{\circ}$, only minor differences are observed on both surfaces. On the upper surface, the effect of the separation correction on the transition location is minimal because transition occurs very close to the leading edge ($x/c < 0.04$). In contrast, flow separation begins much farther downstream (at $\alpha = 14^\circ$, at $x/c = 0.59$ and $x/c = 0.49$ for the SST-$\gamma$ and SST-$\gamma$–$\lambda_{\theta}$ models, respectively), where the separation correction becomes relevant. Similarly, on the lower surface, no significant effects are observed due to the negligible magnitude of the correction. In summary, the separation correction has limited impact on the transition locations of the S809 airfoil. Therefore, the transition model primarily determines the location of transition irrespective to whether or not the proposed turbulence model correction is employed.

Overall, these observations indicated that the proposed separation correction is improving the prediction of lift and drag around the stall angle for both transitional and turbulent cases for the S809 airfoil. Since the model coefficients were calibrated for the S809 case, we now turn to other airfoils to assess the generalizability of the model across geometries, Reynolds numbers, and freestream turbulence intensities.

\begin{figure}[h]
	\centering
	\floatsetup[figure]{style=plaintop} 
	\subcaptionbox{}{\includegraphics[width=0.49\textwidth]{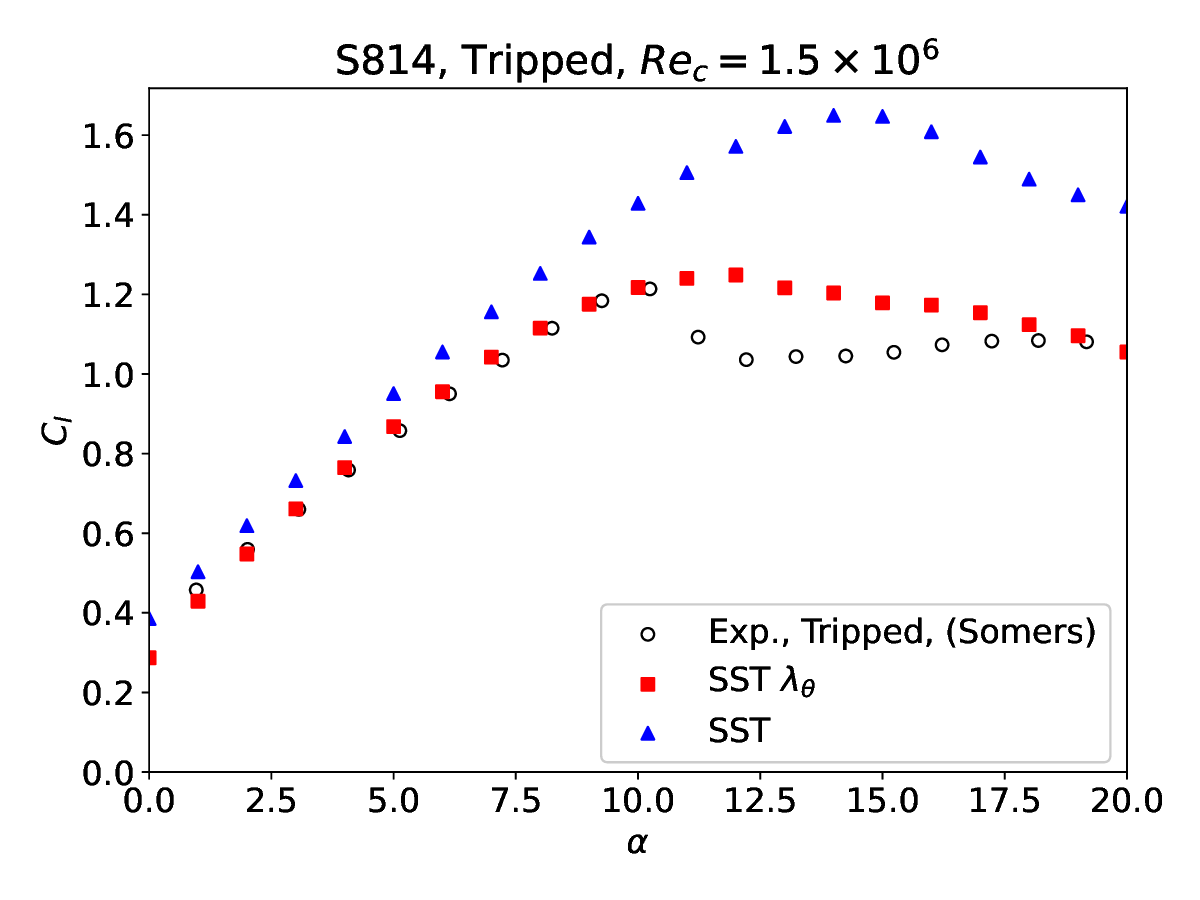}}
	\subcaptionbox{}{\includegraphics[width=0.49\textwidth]{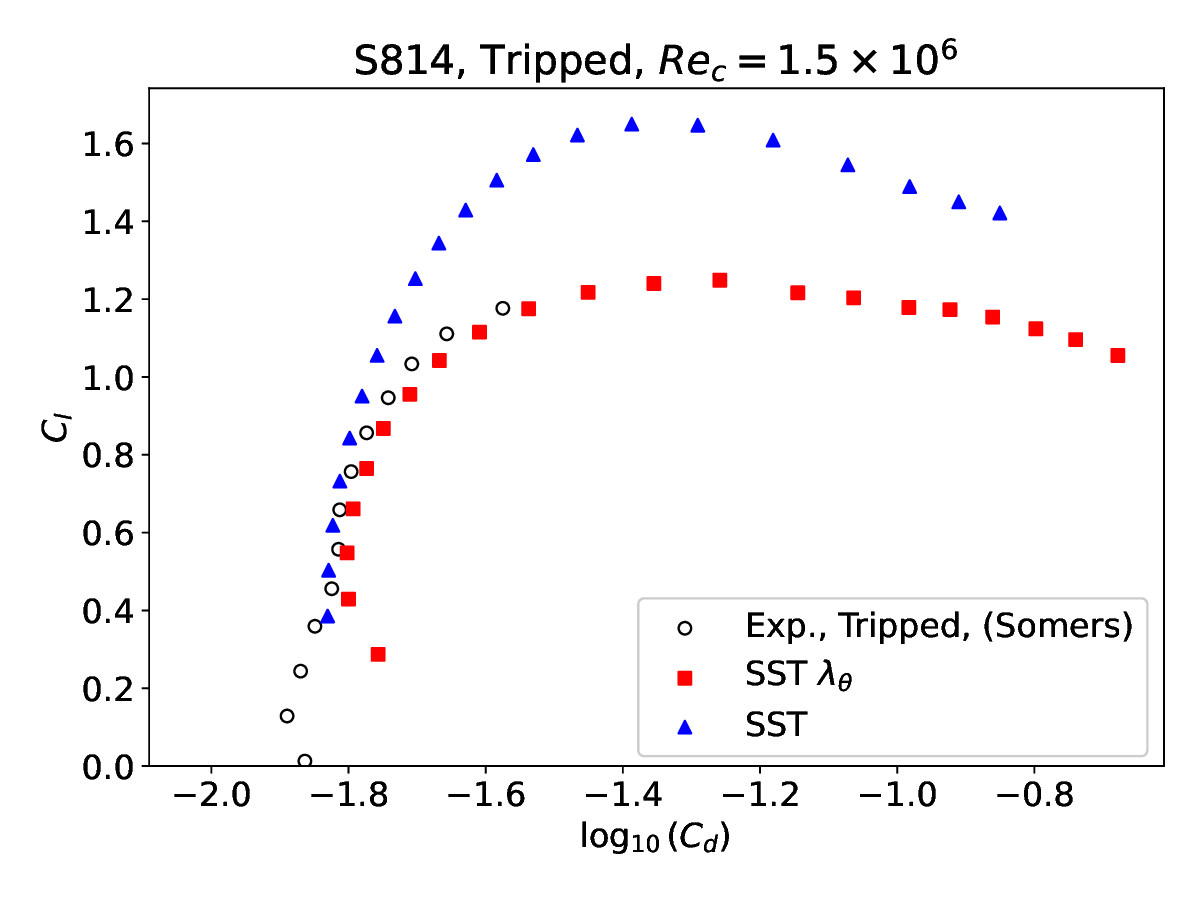}}
	\subcaptionbox{}{\includegraphics[width=0.49\textwidth]{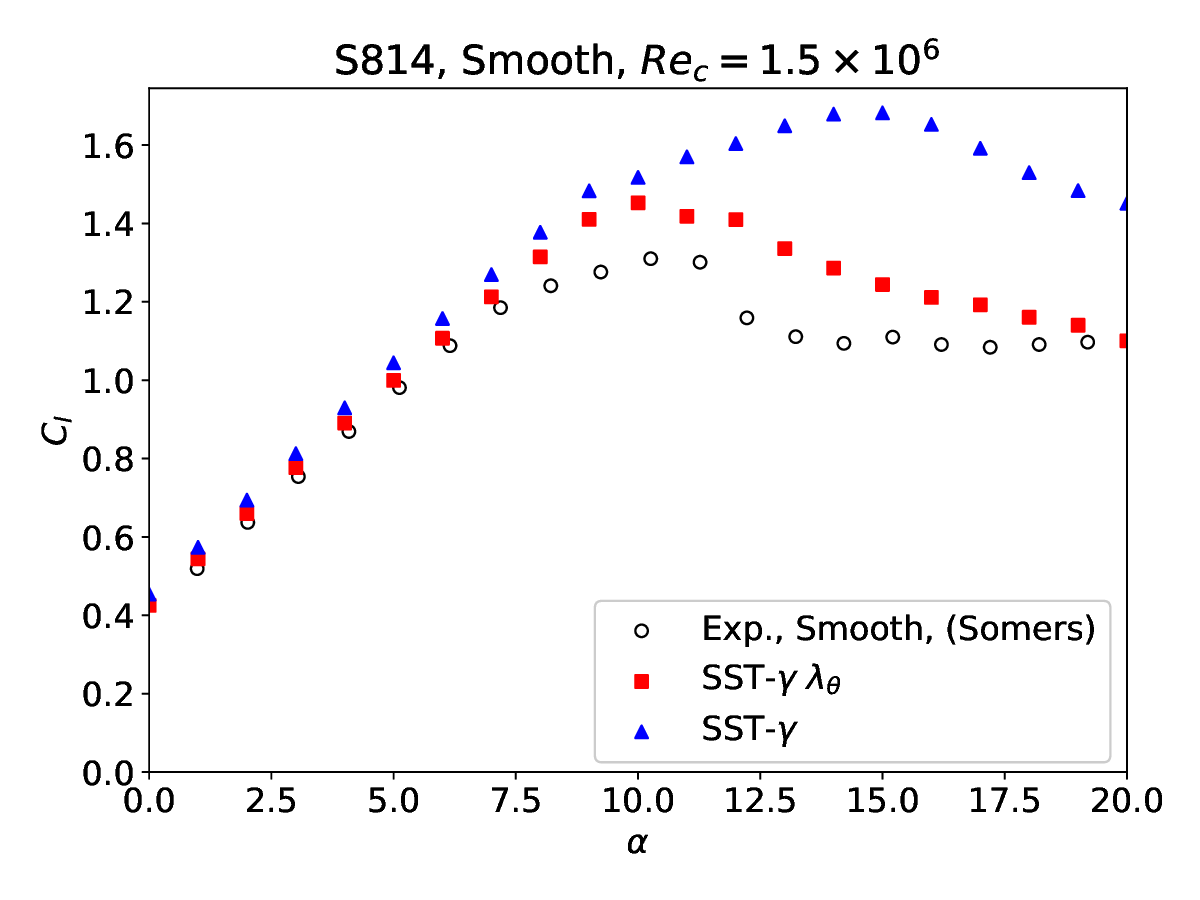}}
	\subcaptionbox{}{\includegraphics[width=0.49\textwidth]{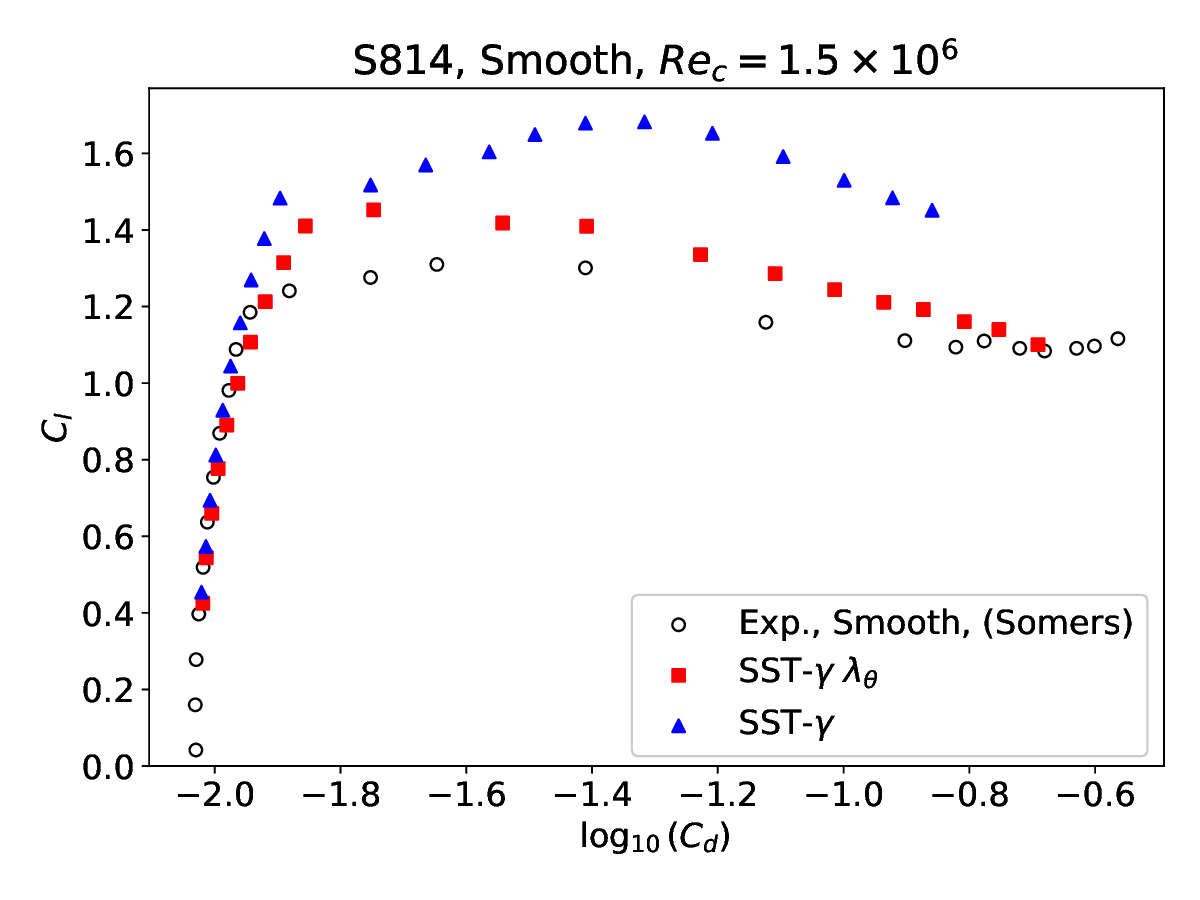}}
	\caption{Coefficient of lift versus (a,c) angle of attack and (b,d) coefficient of drag for the S814 airfoil for the (a,b) tripped and (c,d) smooth cases. The results for the proposed model (red) and baseline model (blue) are compared with experimental reference data (black) \citep{Somers1997_s814}.}
	\label{fig:s814}
\end{figure}

We begin our study of generalization with the thickest of the airfoils considered, the S814, which has a maximum thickness of 24\% of the chord length. The Reynolds number is slightly lower than for the S809 and the turbulence intensity is held constant (see Table \ref{tab:airfoils}). For the S814 airfoil, in Figure \ref{fig:s814}, the lift and drag predictions are improved when the $\lambda_\theta$ correction is applied to the SST model. This observation holds both for transitional (smooth) and fully turbulent (tripped) cases. Unlike for the S809, for the S814, drag data is available at higher angles of attack for the smooth case and the proposed model agrees well with this data suggesting the correction is improving drag predictions well past the onset of stall. Remarkably, for this thick airfoil, the baseline SST model fails to accurately predict lift in the linear regime of the lift vs angle of attack curve ($\alpha \lessapprox 7$). The proposed separation correction resolves this issue. This is consistent with the observations by Griffin et al. \citep{griffin2025improved} that for thick, highly cambered airfoils, separation on the pressure side of the airfoil can occur at low angles of attack manifesting as a vertical shift of the linear region of the lift versus angle of attack curve. The fact that the proposed separation correction accurately predicts pressure-side separation builds confidence in deploying the model on other thick airfoils.

\begin{figure}[h]
	\centering
	\floatsetup[figure]{style=plaintop} 
	\subcaptionbox{}{\includegraphics[width=0.49\textwidth]{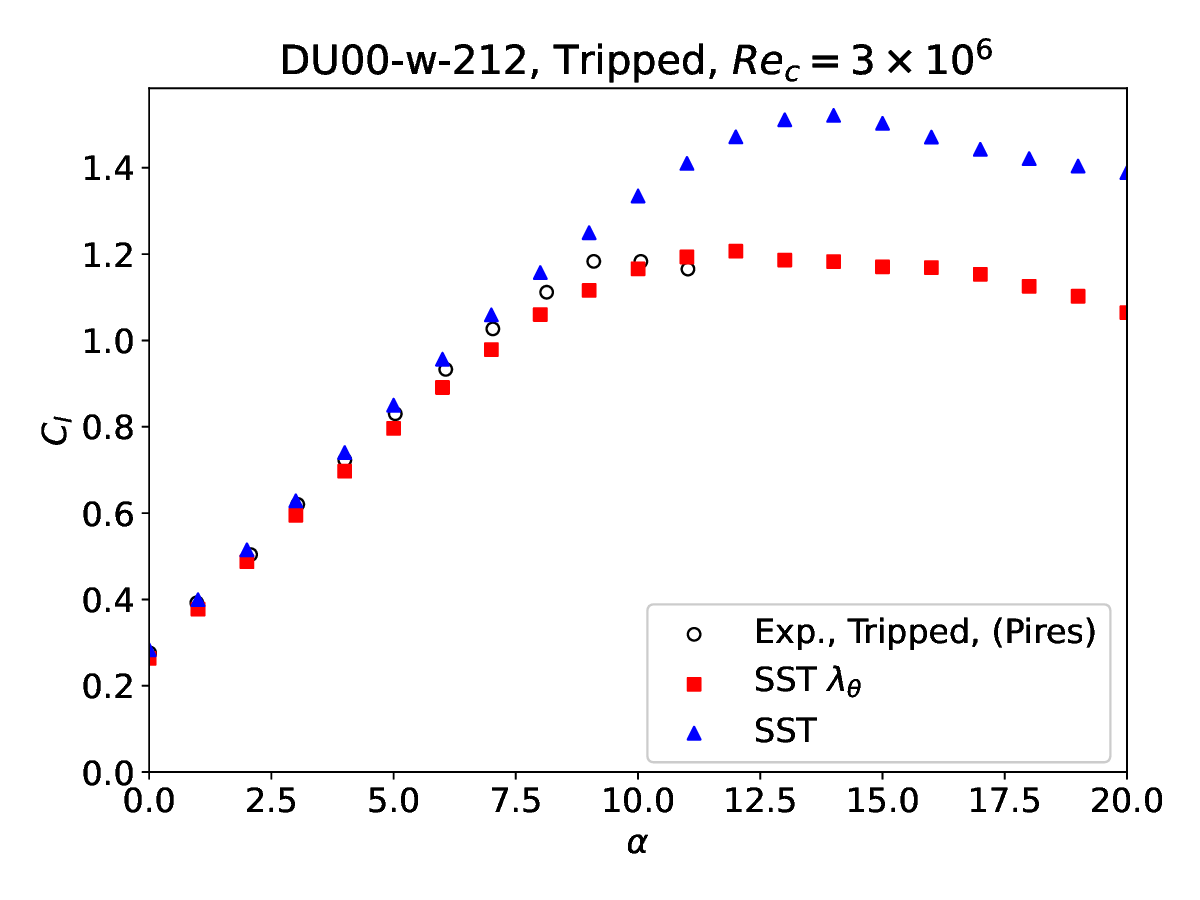}}
	\subcaptionbox{}{\includegraphics[width=0.49\textwidth]{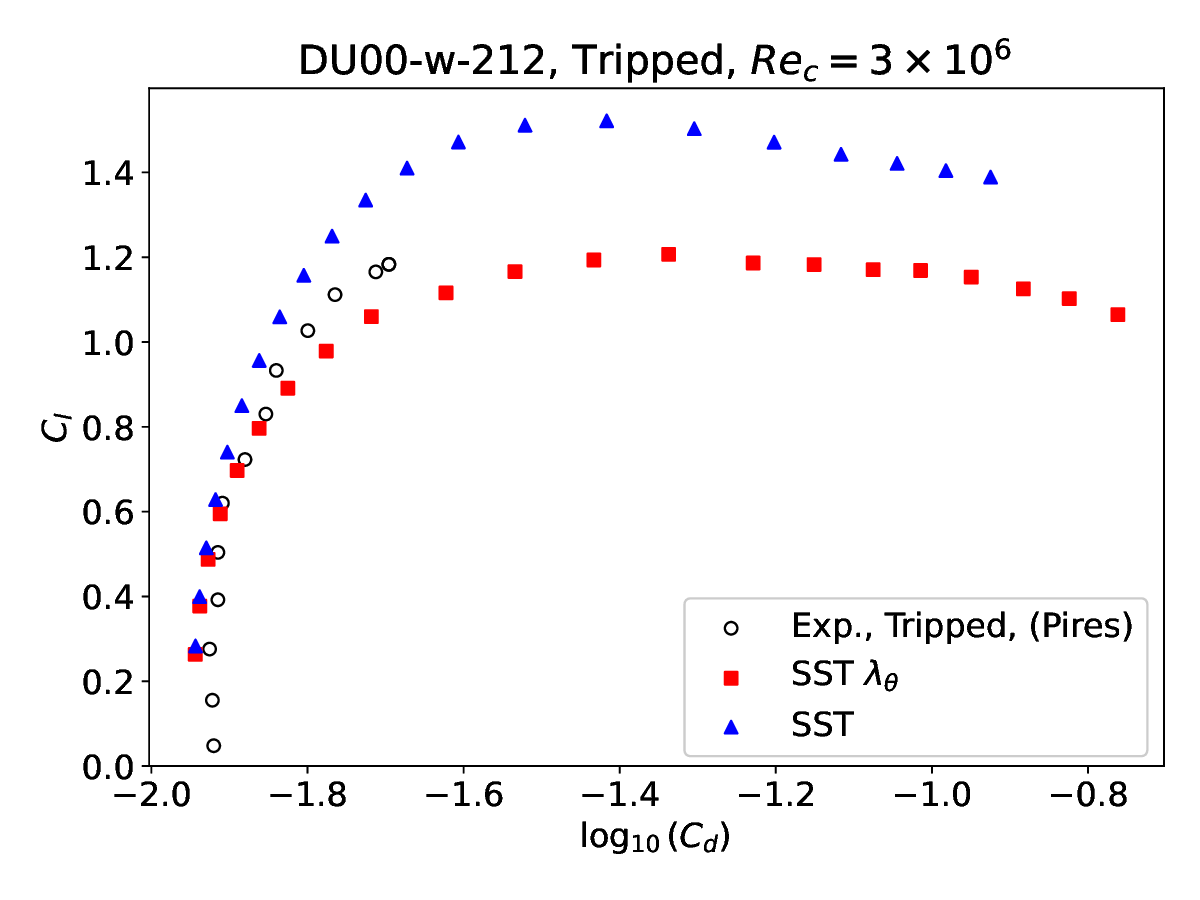}}
	\subcaptionbox{}{\includegraphics[width=0.49\textwidth]{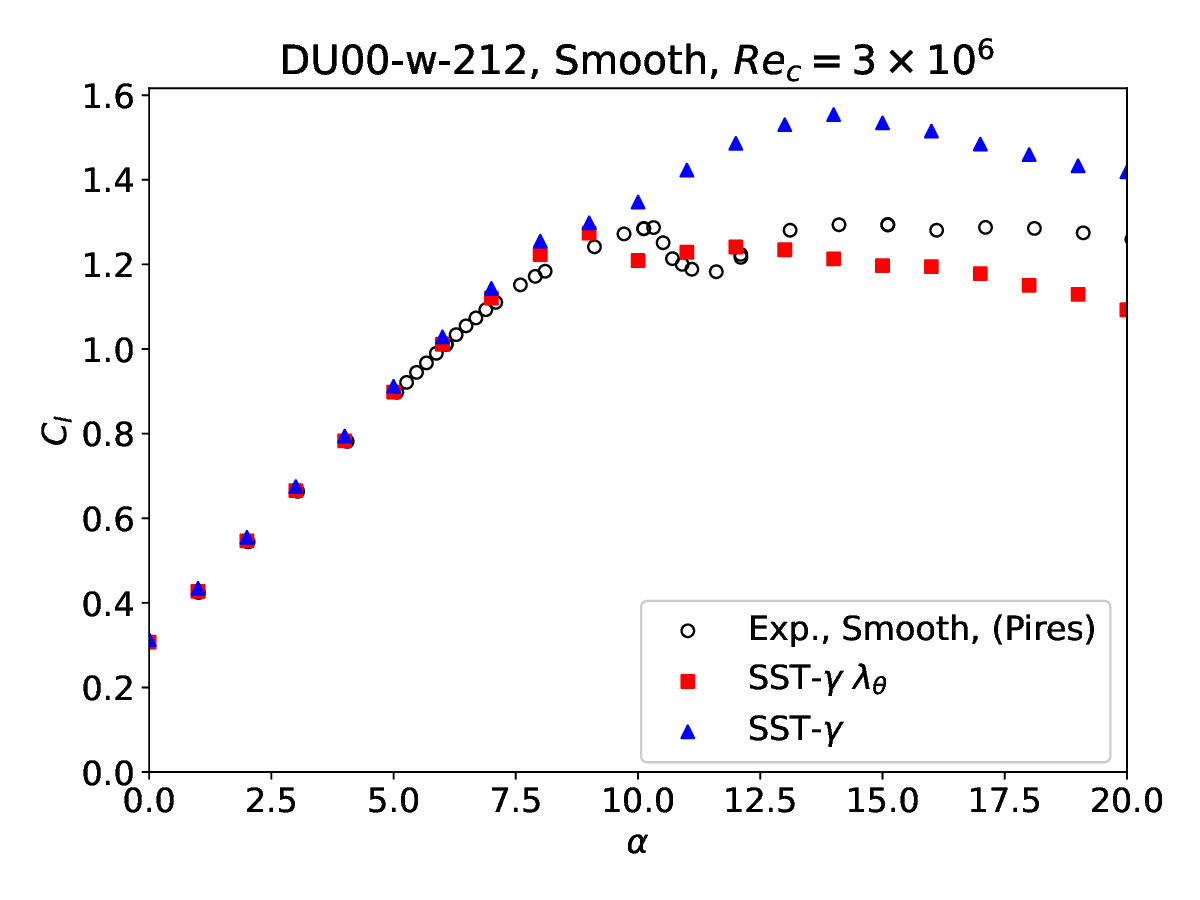}}
	\subcaptionbox{}{\includegraphics[width=0.49\textwidth]{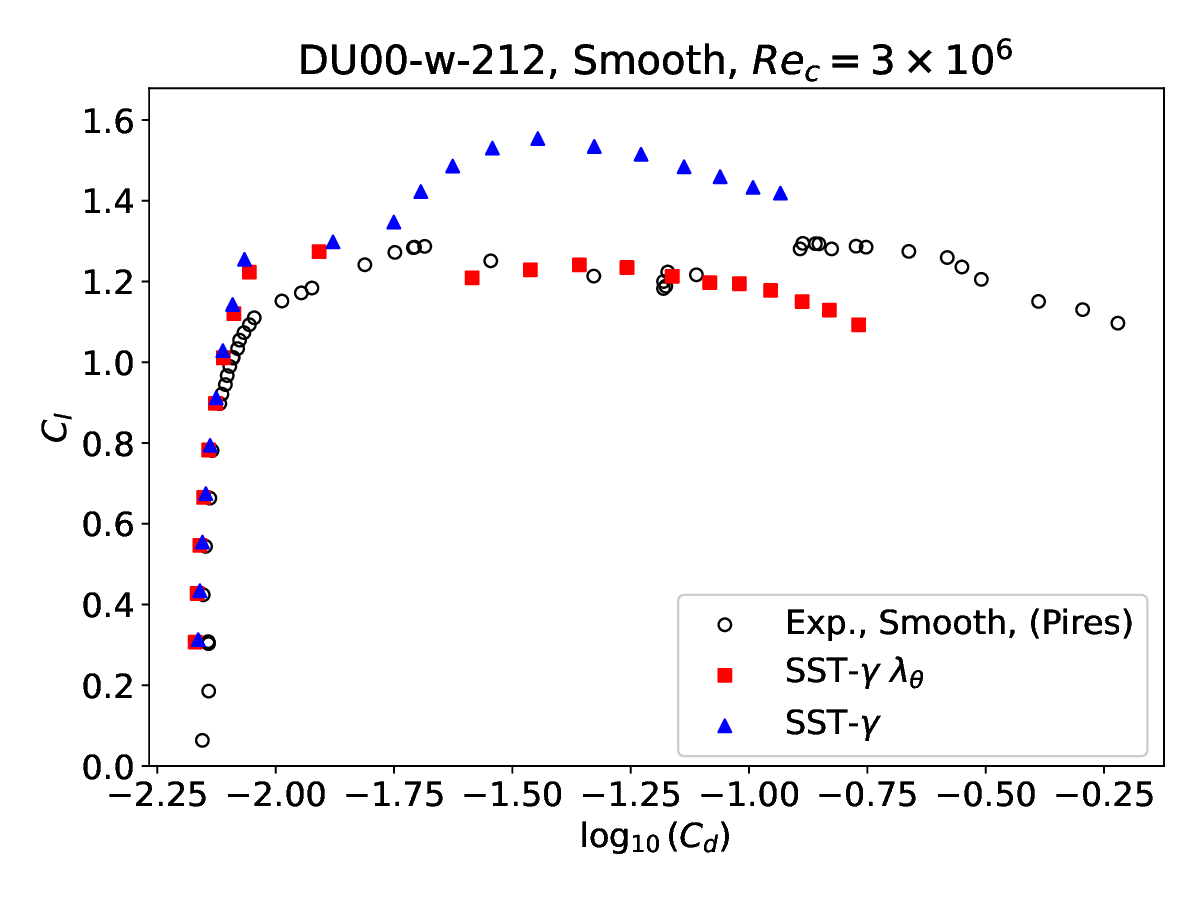}}
	\caption{Coefficient of lift versus (a,c) angle of attack and (b,d) coefficient of drag for the DU00-w-212 airfoil for the (a,b) tripped and (c,d) smooth cases. The results for the proposed model (red) and baseline model (blue) are compared with experimental reference data (black) \citep{Pires2016}.}
	\label{fig:du00w212_R03M}
\end{figure}

The next airfoil considered is the DU00-w-212, another wind-turbine airfoil, with a 21\% thickness and a higher turbulence intensity (see Table \ref{tab:airfoils}). In Figure \ref{fig:du00w212_R03M}, the prediction of the lift coefficient improves when the separation correction is applied, especially near the point of stall. At very high angles of attack ($\alpha \gtrapprox 12$), neither model is accurate but this is likely due to three-dimensional effects \cite{Bidadi2023} and not of primary interest in this study of two-dimensional \gls{rans} simulations which are targeted at predicting stall onset. For the smooth case, the drag results are greatly improved using the correction except at $\alpha \gtrapprox 12$ where neither model is accurate. For the tripped case, the drag is overpredicted with the correction by about as much as it is underpredicted without the correction. The study that collected the tripped experimental reference data acknowledges that there is significant sensitivity of their results depending on which of two tripping strategies is employed \cite{Pires2016}, which undermines our assumption that their results represent a fully turbulent boundary layer. As a result, it is unclear what the effect of our separation correction is on the accuracy of drag predictions for the fully turbulent simulations.

\begin{figure}[h]
	\centering
	\floatsetup[figure]{style=plaintop} 
	\subcaptionbox{Pressure coefficient}{\includegraphics[width=0.49\textwidth]{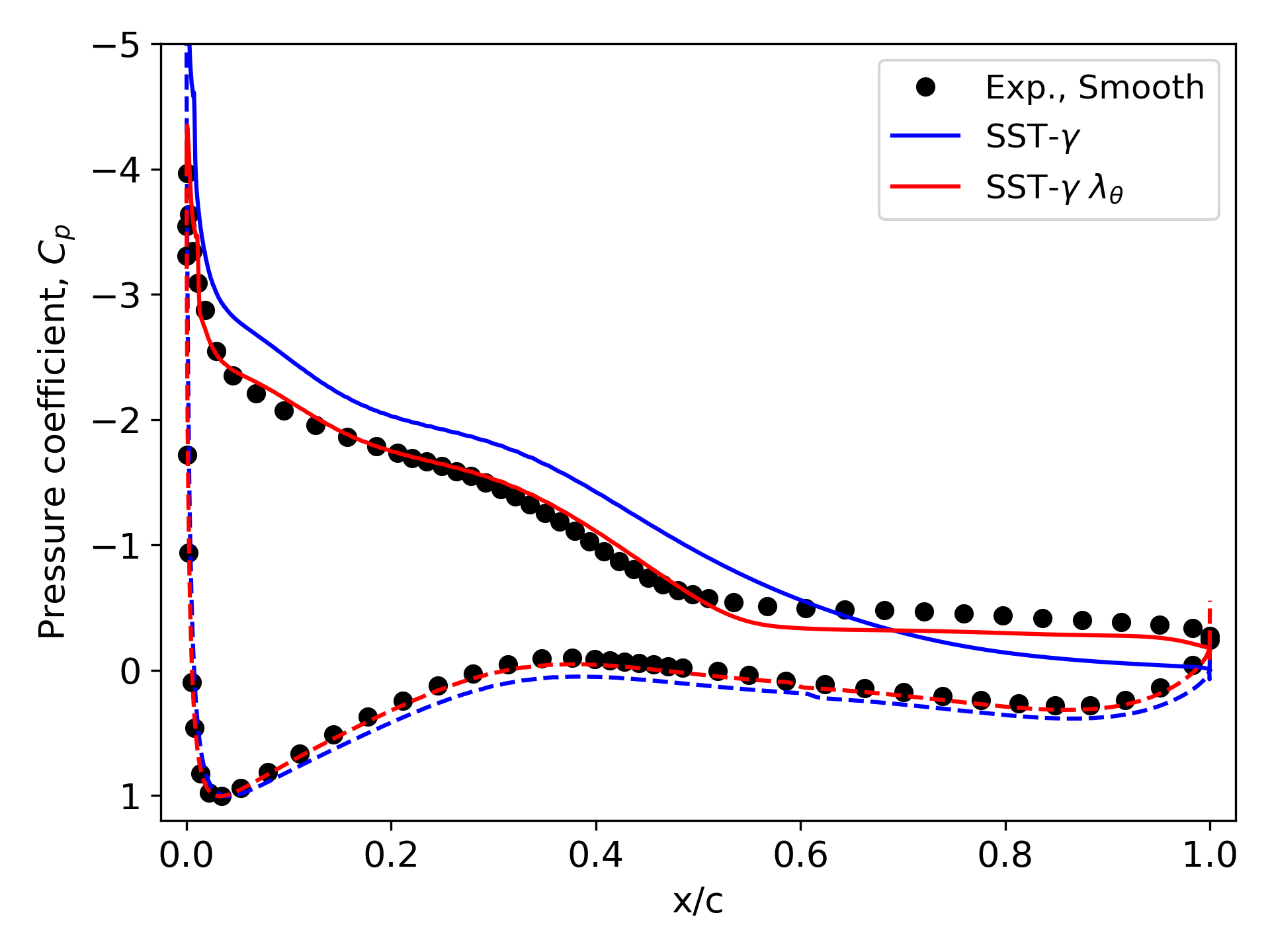}}
	\subcaptionbox{Skin friction coefficient}{\includegraphics[width=0.49\textwidth]{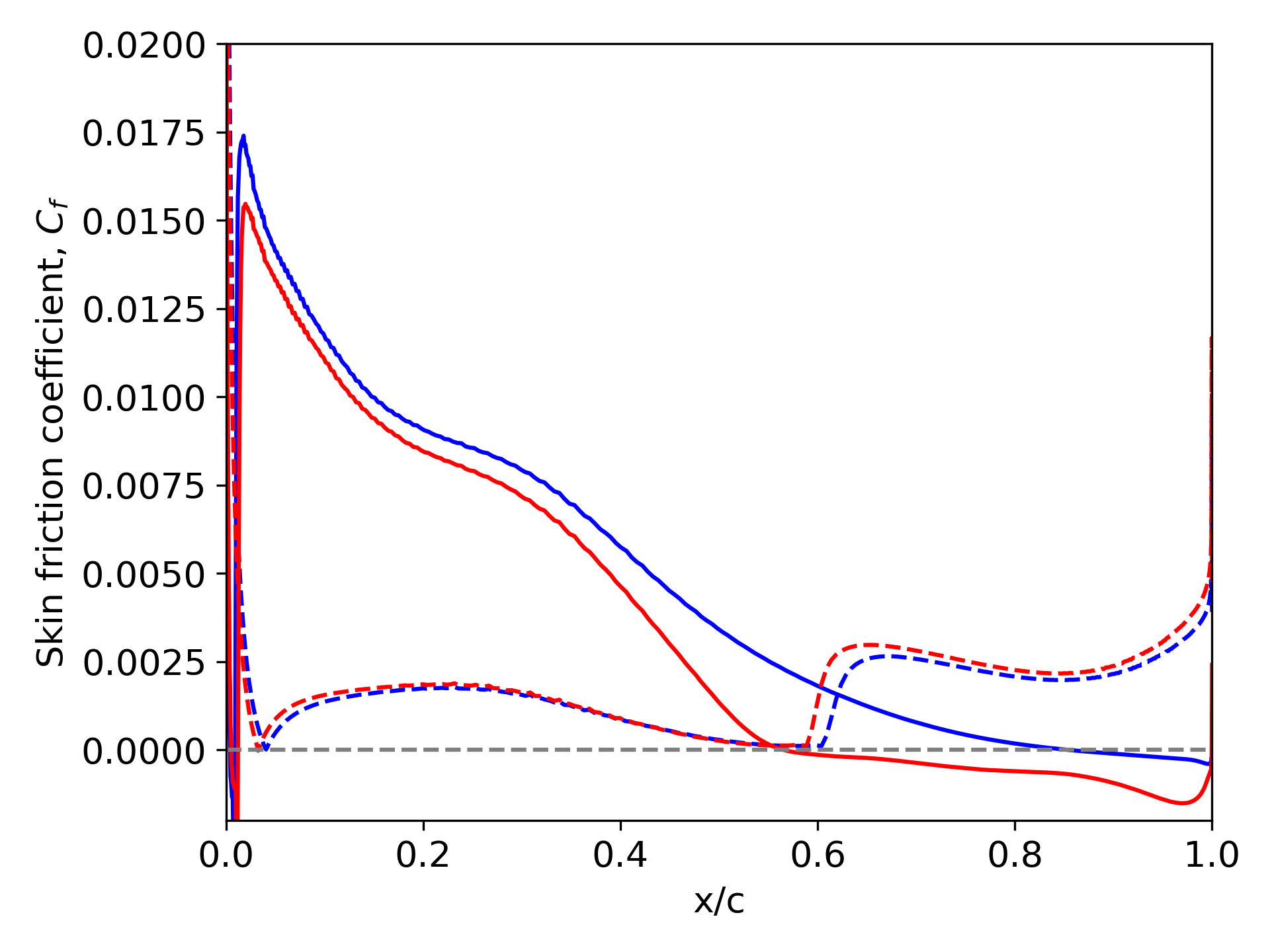}}
	\caption{Comparison of the (a) pressure and (b) skin friction coefficients at $\alpha=12^{\circ}$ for the smooth DU00-w-212 airfoil. Solid/Dashed line: Upper/Lower surfaces, respectively.}
	\label{fig:du00w212_cpcf}
\end{figure}

Figure~\ref{fig:du00w212_cpcf} presents comparisons of the pressure and skin friction coefficients for the smooth case at $\alpha = 12^{\circ}$, with and without the separation correction. For the pressure coefficient, the available experimental data are also included for reference. The corrected model shows a markedly improved agreement with the experimental pressure distribution, whereas the uncorrected model fails to capture the overall trend on the upper surface (solid line).

The skin friction coefficient distributions further illustrate the impact of the correction. With the correction, flow separation on the upper surface occurs at approximately $x/c = 0.566$, compared to $x/c = 0.854$ without the correction. On the lower surface, the corrected model also exhibits improved behavior, showing a slightly earlier transition location than the uncorrected result.

\begin{figure}[h]
	\centering
	\floatsetup[figure]{style=plaintop} 
	\subcaptionbox{}{\includegraphics[width=0.49\textwidth]{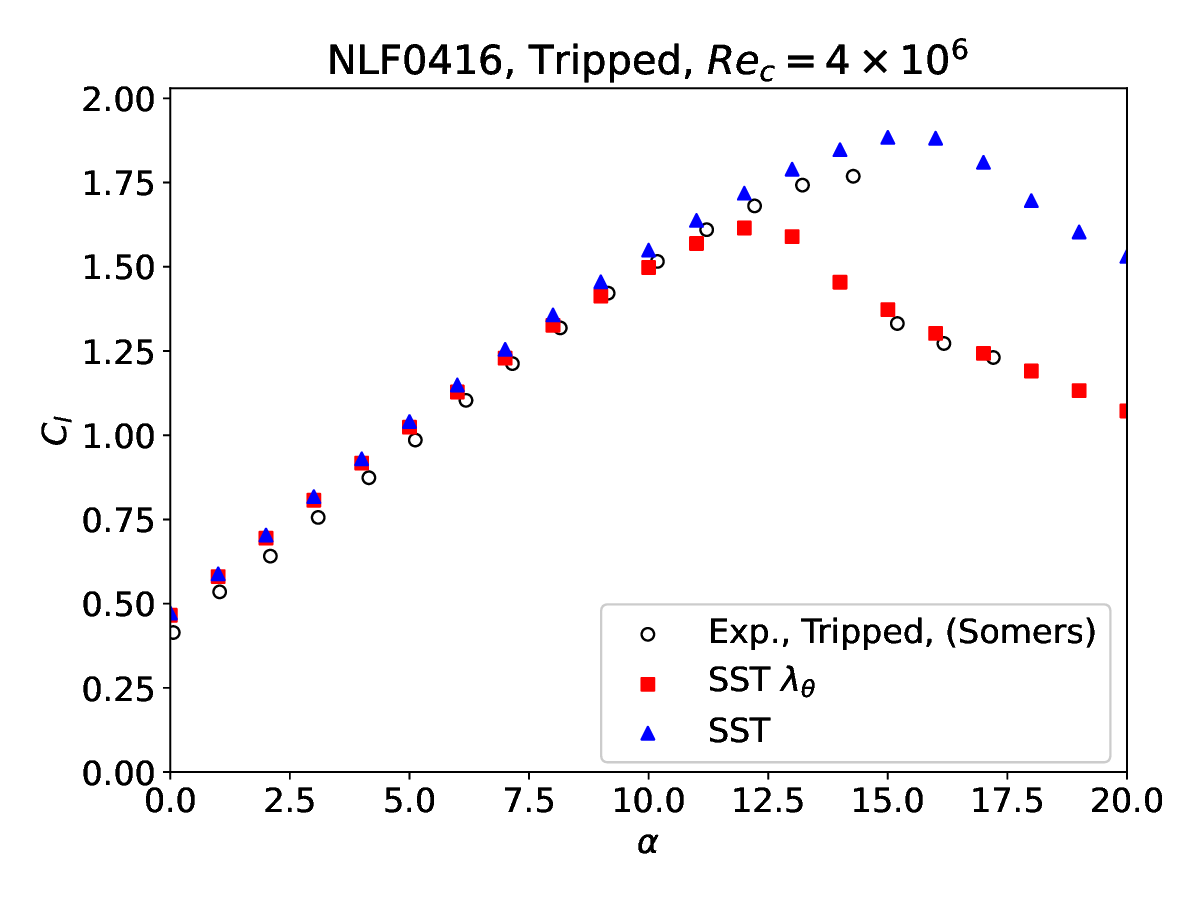}}
	\subcaptionbox{}{\includegraphics[width=0.49\textwidth]{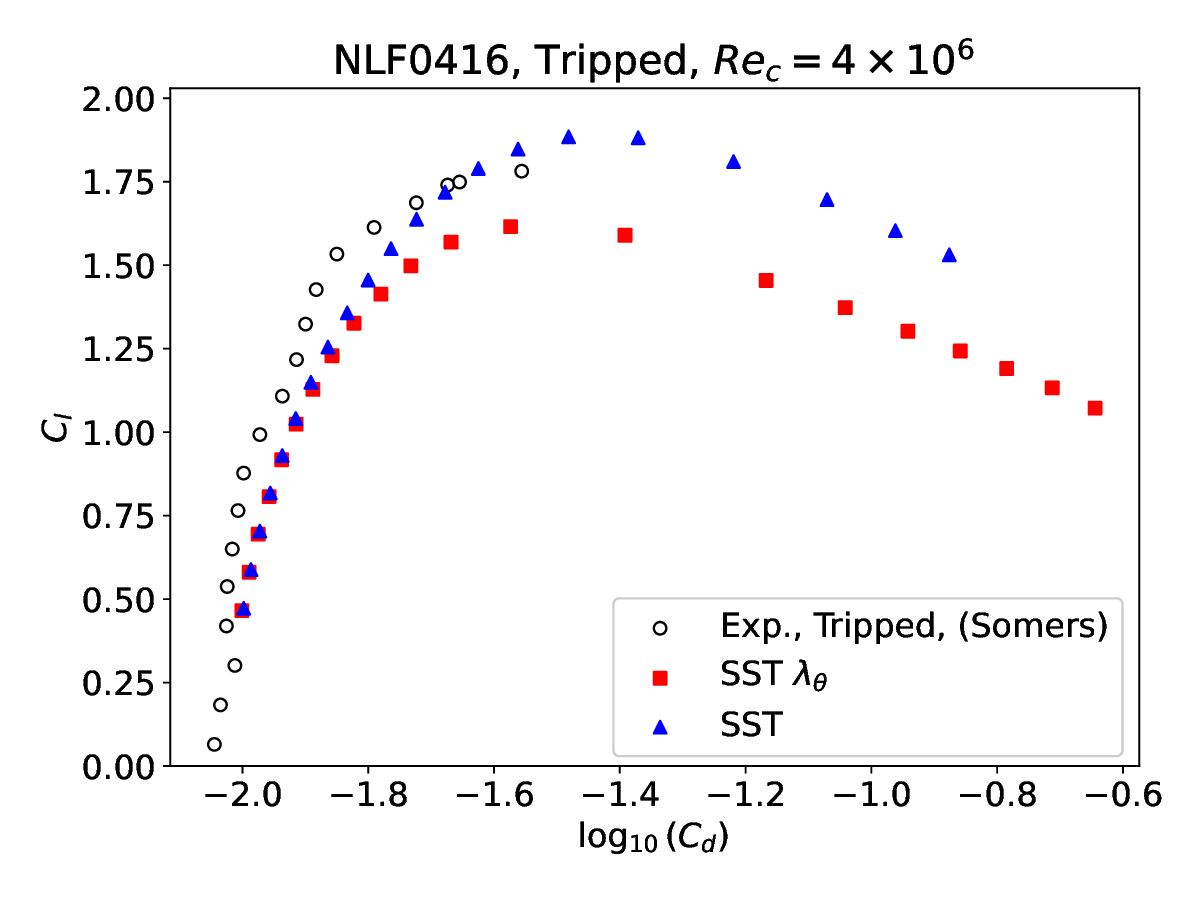}}
    \subcaptionbox{}{\includegraphics[width=0.49\textwidth]{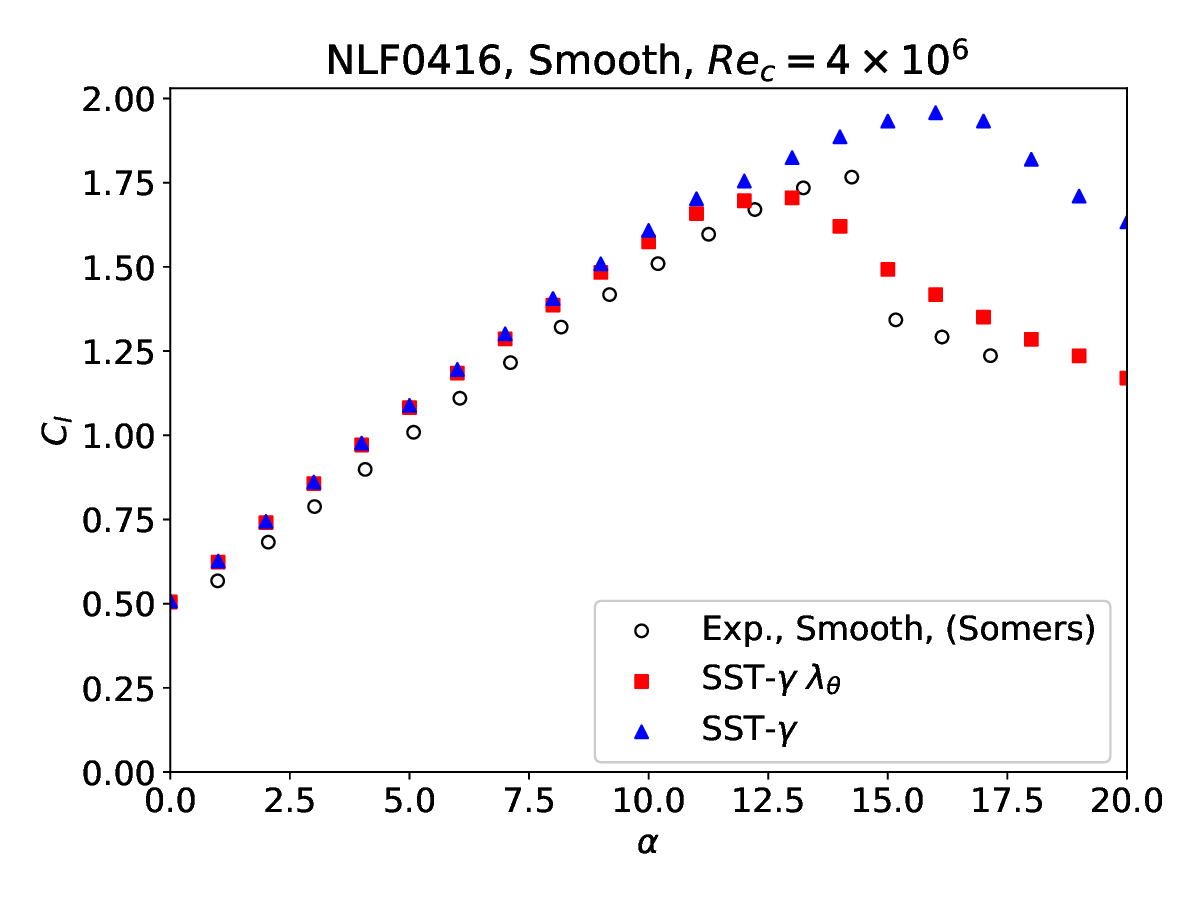}}
	\subcaptionbox{}{\includegraphics[width=0.49\textwidth]{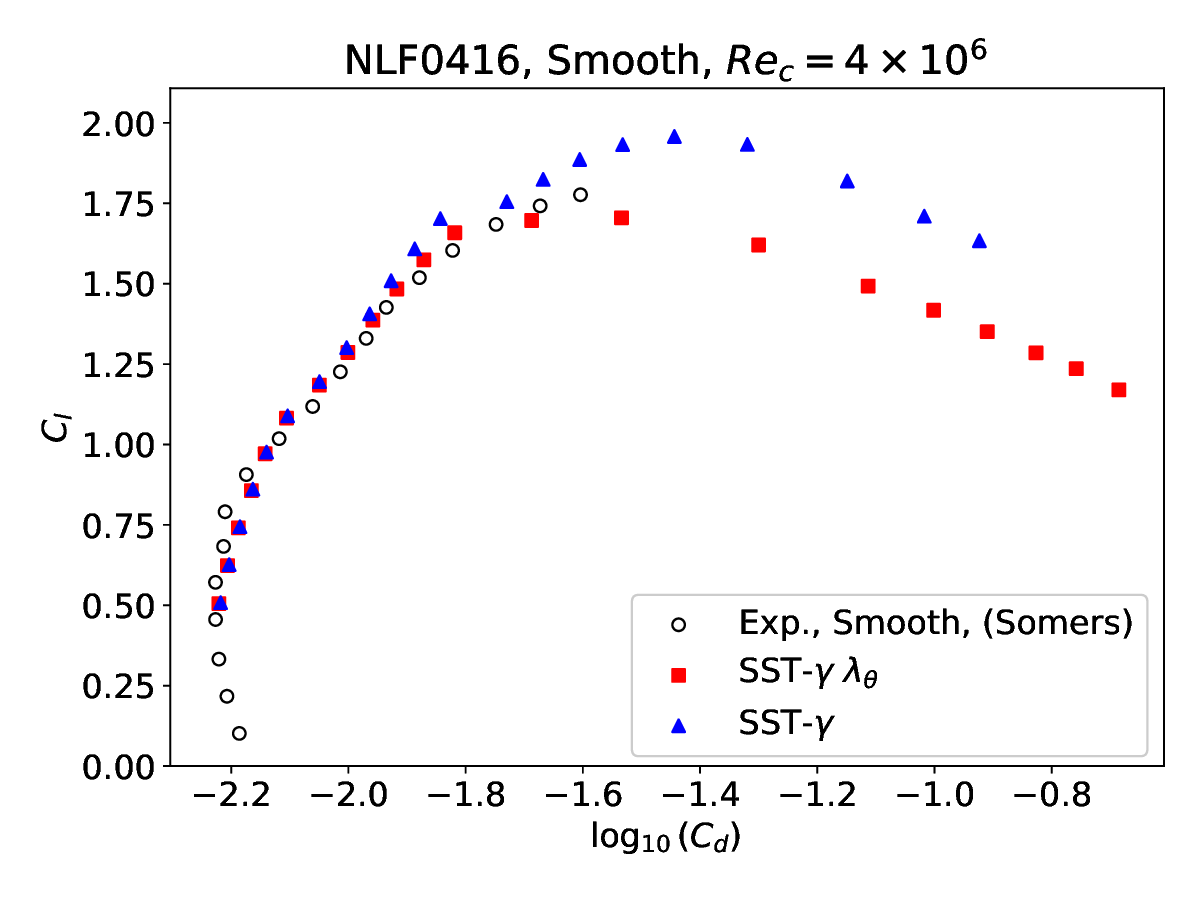}}
	\caption{Coefficient of lift versus (a,c) angle of attack and (b,d) coefficient of drag for the NLF(1)-0416 airfoil for the (a,b) tripped and (c,d) smooth cases. The results for the proposed model (red) and baseline model (blue) are compared with experimental reference data (black) \citep{Somers1981}.}
	\label{fig:nlf0416}
\end{figure}

Turning now to thinner airfoils representative of aerospace applications, the natural laminar flow (NLF)(1)-0416 is considered. This airfoil has 16\% thickness and has a unique profile that delays the onset of turbulent transition, making it an interesting challenge for the transition and separation models. In Figure \ref{fig:nlf0416}(a,c), the experimental data exhibits a sharp drop in lift beyond the stall angle, which is characteristic of thinner airfoils. Neither the baseline nor proposed models capture the sharp drop, but the proposed model with and without transition predict stall about two degrees early, which leads to a significantly more accurate lift predictions post stall $\alpha> 15$. Both models slightly overpredict lift in the linear region ($\alpha \lessapprox 11$) particularly for the smooth case. Drag data is not collected past the point of stall, which limits the conclusions that can be drawn about the effect of the correction on the drag prediction. However, for the tripped case, both models underpredict drag for the low angles of attack perhaps due to the unique design of the airfoil designed to delay transition. The underprediction of the proposed model is even larger at higher angles of attack due to the early prediction of stall. For the smooth case, both models accurately predict drag across the range of angles of attack where experimental data are available.

\begin{figure}[h]
	\centering
	\floatsetup[figure]{style=plaintop} 
	\subcaptionbox{Pressure coefficient}{\includegraphics[width=0.49\textwidth]{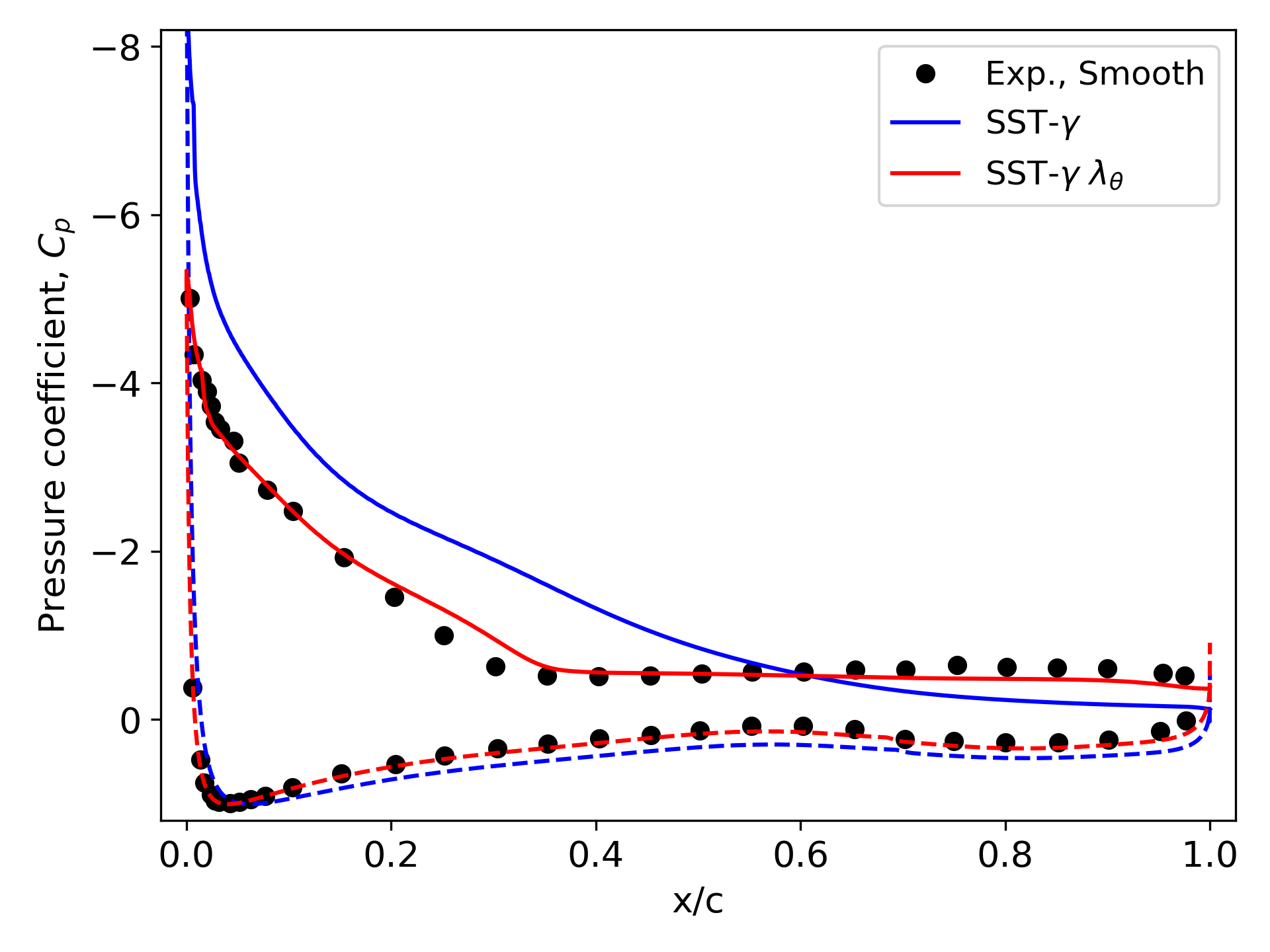}}
	\subcaptionbox{Skin friction coefficient}{\includegraphics[width=0.49\textwidth]{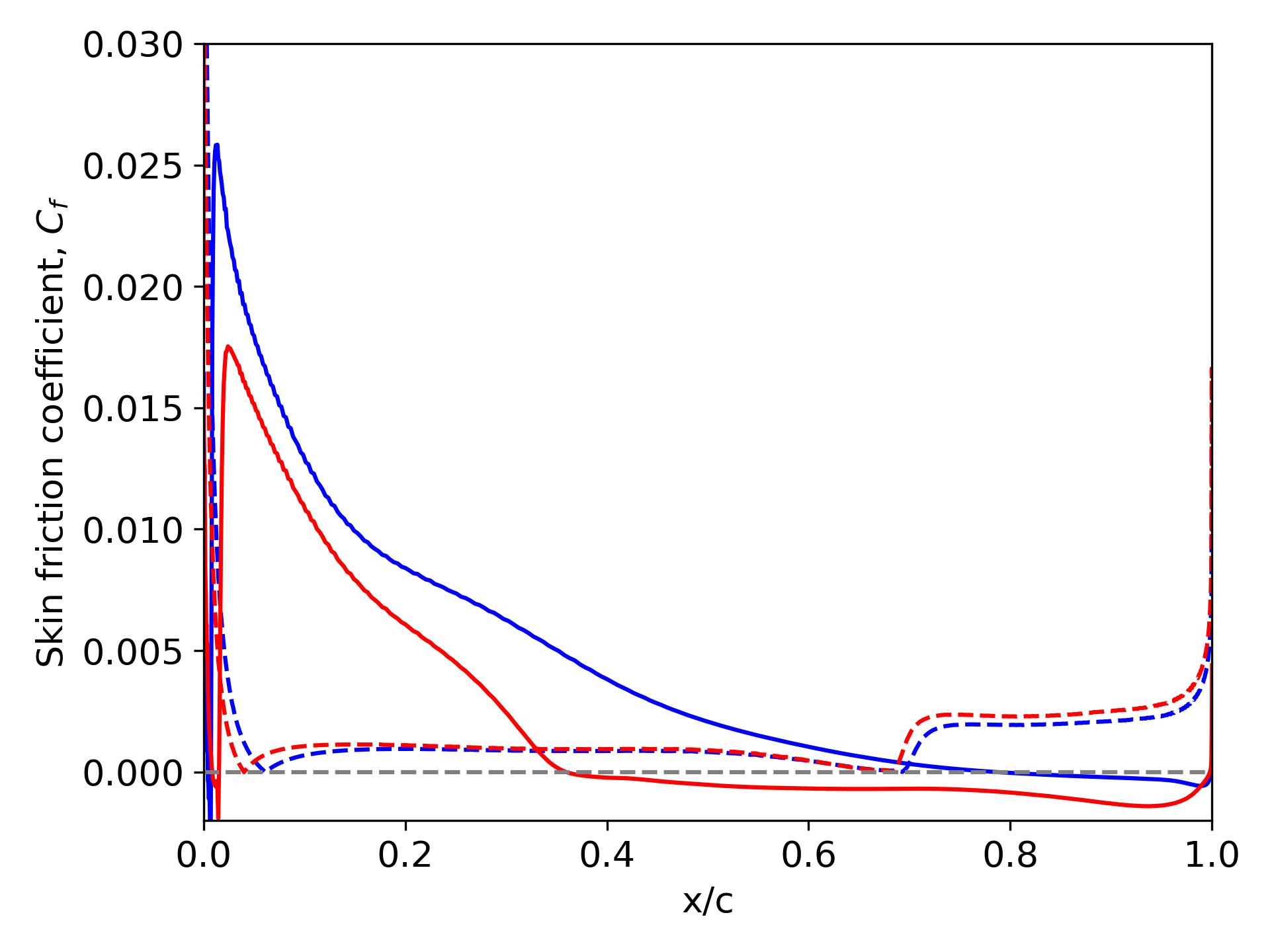}}
	\caption{Comparison of the (a) pressure and (b) skin friction coefficients at $\alpha=16^{\circ}$ for the smooth NLF(1)-0416 airfoil. Solid/Dashed line: Upper/Lower surfaces, respectively.}
	\label{fig:NLF-0416_cpcf}
\end{figure}

Figure~\ref{fig:NLF-0416_cpcf} presents the pressure and skin friction coefficients for the smooth case at $\alpha = 16^{\circ}$. At this angle of attack, both simulations overpredict the lift coefficient. However, the corrected model yields a substantially reduced post-stall lift compared with the uncorrected model. This trend is also reflected in the pressure distributions. Without the correction, the simulation predicts excessively high suction pressure on the upper surface up to $x/c = 0.6$, indicating delayed separation. In contrast, the corrected model shows a flattened pressure distribution beginning at $x/c = 0.36$ due to earlier separation, resulting in significantly improved agreement with the experimental data.
The skin friction coefficient distributions also confirm this behavior. On the upper surface, the predicted separation location shifts markedly toward the leading edge when the correction is applied, moving from $x/c = 0.785$ in the original model to $x/c = 0.36$ with the separation correction.

The last airfoil considered is the NACA23012, which is 12\% thick and more characteristic of the cross section of an aircraft wing. Forced transition data is unavailable from this experiment \cite{Broeren2018} as it is primarily an investigation of aircraft icing. As the thinnest airfoil considered, the NACA23012 also has the sharpest drop in lift after the stall angle, a phenomenon that the separation correction captures well leading to significantly improved lift predictions compared to the baseline model. Drag is underpredicted at low angles of attack despite the fact that lift is accurately predicted. This suggests that the source of the discrepancy may be the transition model. The experimental investigation does not quantify the turbulence intensity \cite{Broeren2018}, but prior investigations in the University of Illinois at Urbana-Champaign low-speed, low-turbulence wind tunnel report a turbulence intensity below 0.1\% \cite{Selig2004}. The simulations were performed setting the intensity to 0.1\%, but aberrations from this value are likely and the transition model is sensitive to small changes in the freestream turbulence intensity. This precludes quantitative evaluations of the effect of the separation correction to the drag predictions for this case.

\begin{figure}[h]
	\centering
	\floatsetup[figure]{style=plaintop} 
	\subcaptionbox{}{\includegraphics[width=0.49\textwidth]{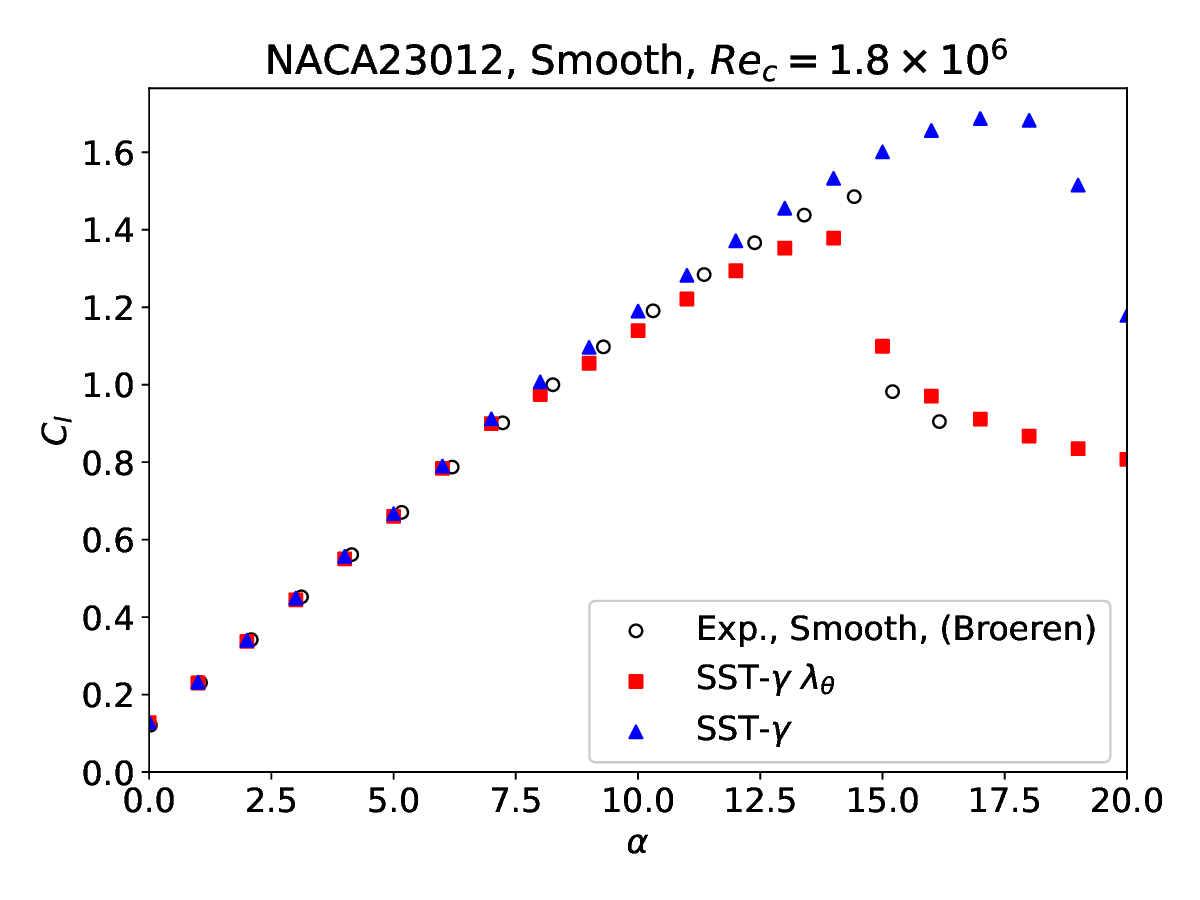}}
	\subcaptionbox{}{\includegraphics[width=0.49\textwidth]{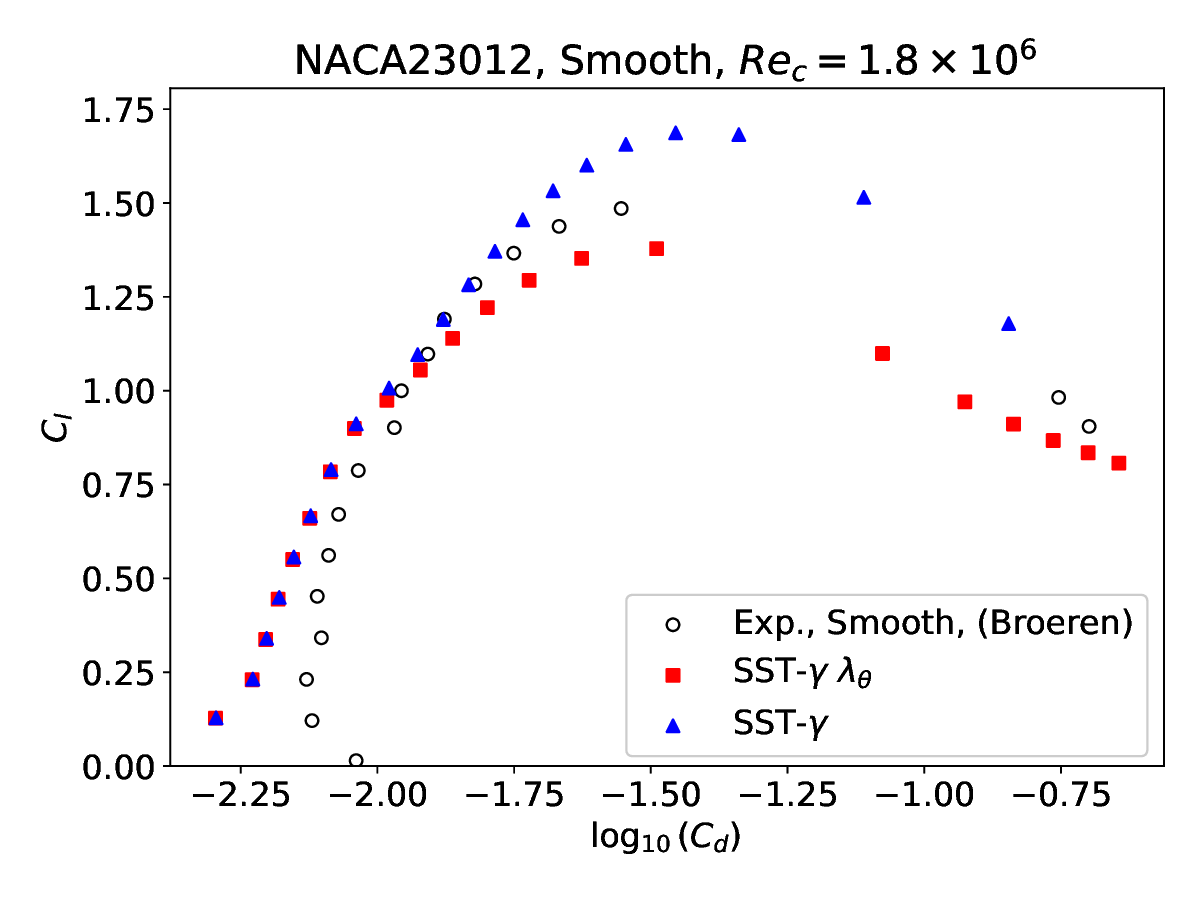}}
	\caption{Coefficient of lift versus (a) angle of attack and (b) coefficient of drag for the NACA23012 airfoil for the smooth case. The results for the proposed model (red) and baseline model (blue) are compared with experimental reference data (black) \citep{Broeren2018}.}
	\label{fig:naca23012}
\end{figure} 

In summary, for the cases study, the model improves the prediction of lift immediately after stall by promoting stall at lower angles of attack than the baseline SST model. This is expected since the proposed model selectively lowers the eddy-viscosity coefficient [see Eq.~(\ref{eq:a1_model})] in regions of strong adverse pressure gradients once the pressure gradient sensor exceeds the critical threshold. Lower eddy viscosity decreases the mixing of high-momentum off-wall fluid with the low-momentum near-wall fluid that is liable to separate. For all cases, except the tripped S814 and the smooth NLF(1)-0416, the baseline SST model is already accurate at low angles of attack and the proposed model does not significantly affect lift or drag in this region. This in and of itself is an accomplishment, since it means the model is not degrading the calibration in attached flows. This was not the case in prior investigations, which showed that direct parameter tuning of the $a_1$ coefficient without the use of a sensor can lead to errors in the prediction of drag on a flat plate \citep{Matyushenko2016}. For the case of the tripped S814 airfoil, the lift is improved even at low angles of attack when the proposed model is used. This may be due to an improved prediction of a pressure-side separation as observed in our prior work \citep{griffin2025improved} with the similar FFA-W3-301 airfoil. For the drag predictions, when the turbulence intensity is precisely quantified experimentally, the proposed correction improves the prediction of drag in the angle of attack regime near stall.

\begin{figure}
	\centering
	\floatsetup[figure]{style=plaintop} 
	\subcaptionbox{}{\includegraphics[width=0.49\textwidth]{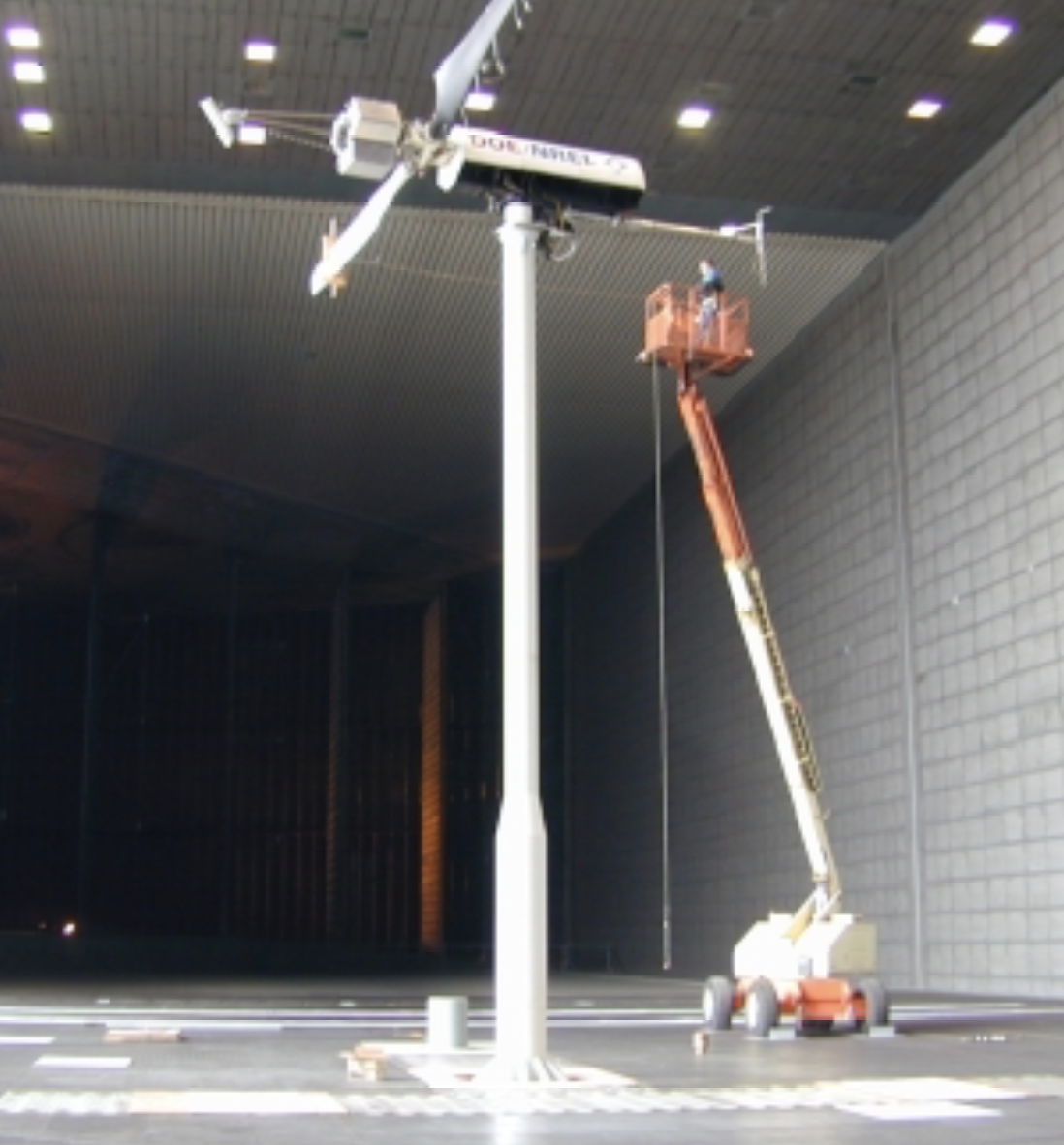}}
	\subcaptionbox{}{\includegraphics[width=0.49\textwidth]{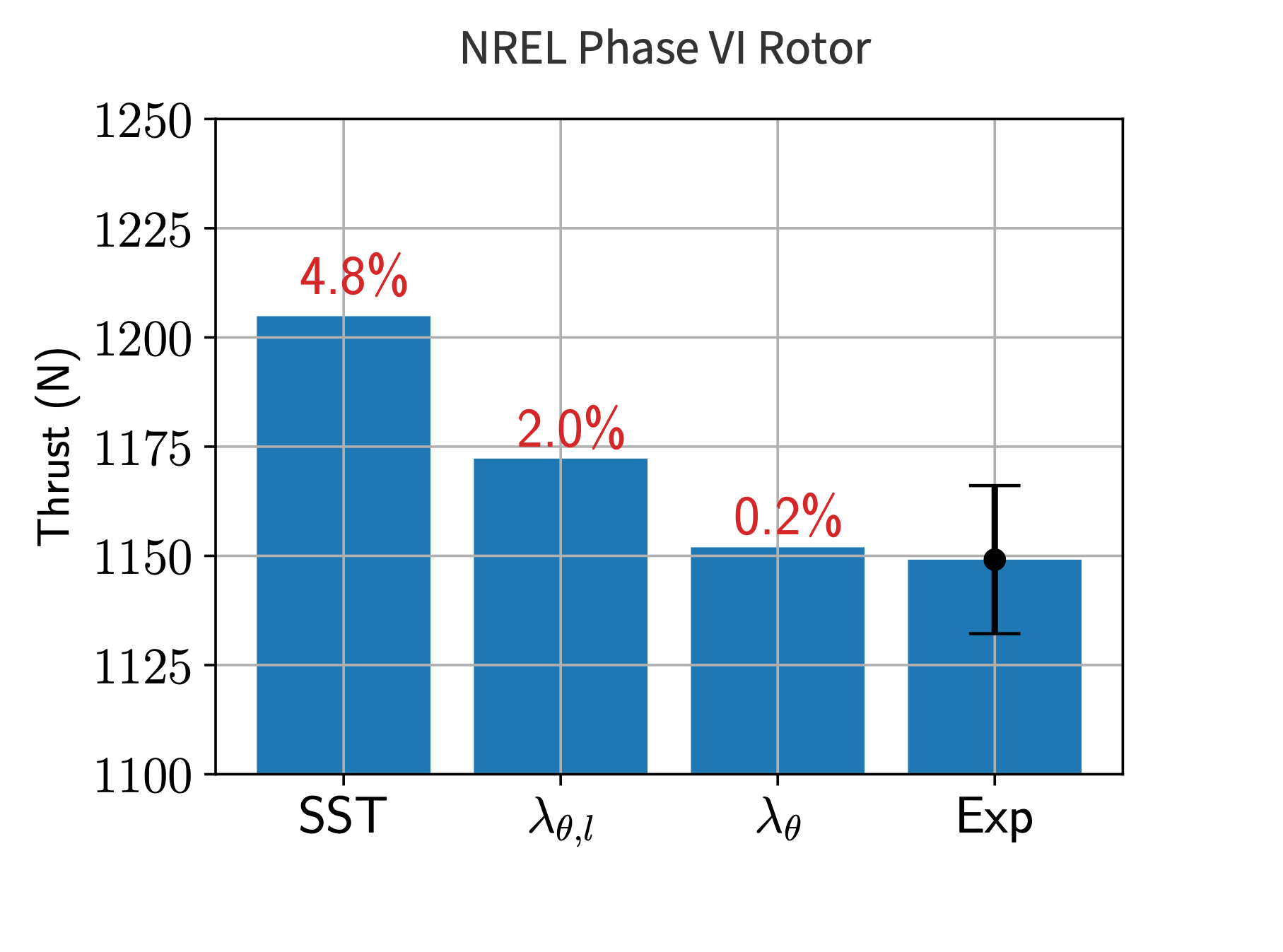}}
	\caption{(a) Image of the wind tunnel test of the NREL Phase VI wind turbine rotor \citep{Hand2001} and (b) corresponding thrust predictions plotted for the baseline model (denoted SST), the laminar separation criterion (denoted $\lambda_{\theta,l}$) and the proposed composite separation criterion (denoted $\lambda_\theta$). The experimental measurements with reported error bars are shown for reference, and the error of the three models is reported in red text.}
	\label{fig:phase_vi}
\end{figure}

To assess the model in a three-dimensional test case, the application to the NREL Phase VI wind turbine rotor is considered. This 10.06 m turbine was experimentally investigated \citep{Hand2001} in a wind tunnel, which makes it a suitable test case for model validation. We consider the case with a freestream velocity of 7 m/s and a frequency of 72 rotations per minute as this corresponds to the onset of separation/stall. At this condition, the baseline SST model and the proposed model predict torque within the experimental uncertainties (not shown). However, the baseline SST overpredicts the thrust by 4.8\% as shown in Figure \ref{fig:phase_vi}. When only the laminar separation criterion $\lambda_{\theta,l}$ is used, the error reduces to 2.0\%. For the proposed model with the composite laminar and turbulent criterion $\lambda_{\theta,c}$ (denoted with $\lambda_\theta$ in all figures for simplicity), the error further reduces to 0.2\%, which is within the range of experimental uncertainty. This indicates that the composite sensor developed in this work is necessary beyond the laminar sensor used previously in transition models. 

\begin{figure}[h]
	\centering
	\floatsetup[figure]{style=plaintop} 
	\subcaptionbox{$r/R=0.47$}{\includegraphics[width=0.49\textwidth]{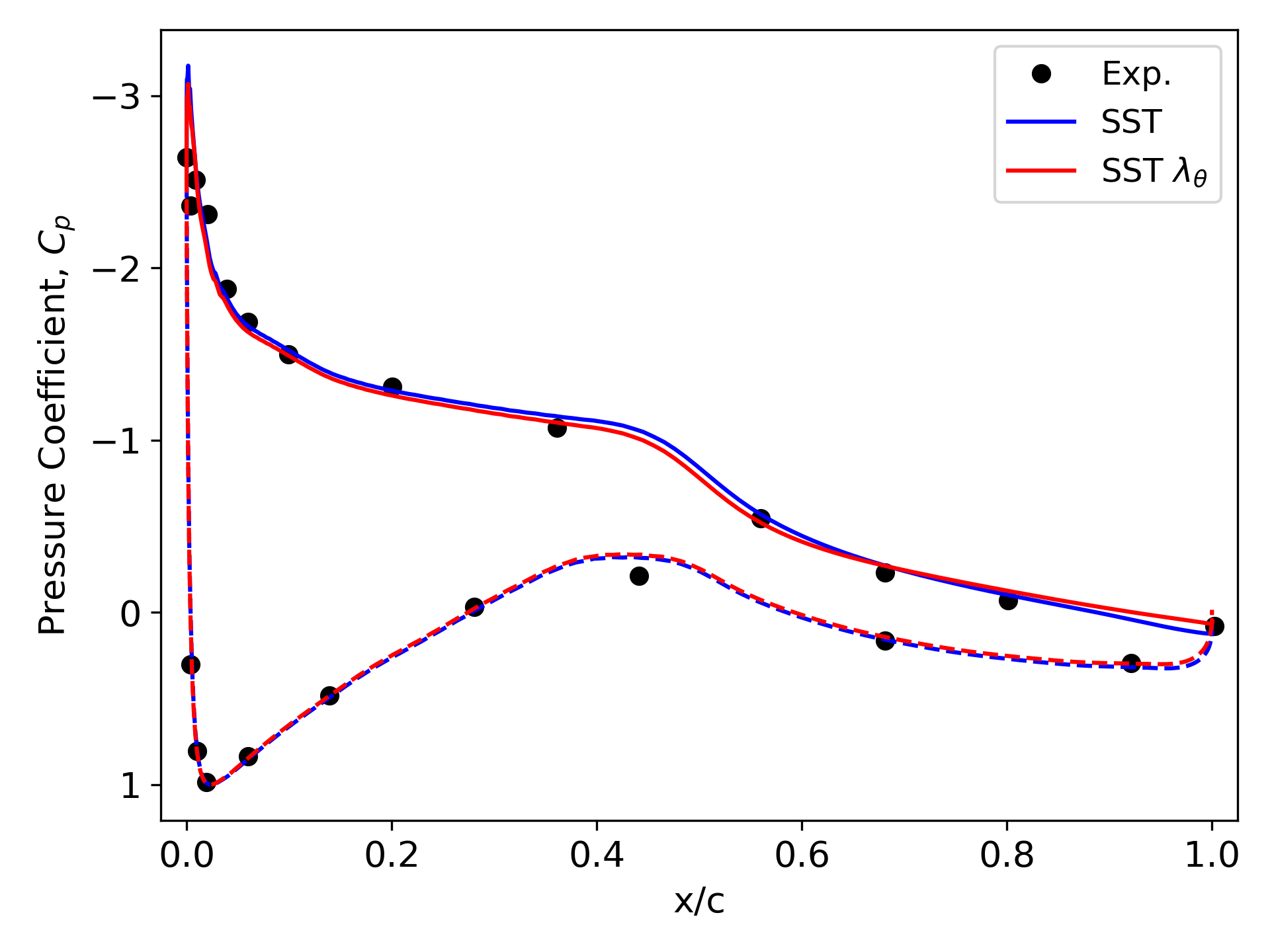}}
	\subcaptionbox{$r/R=0.63$}{\includegraphics[width=0.49\textwidth]{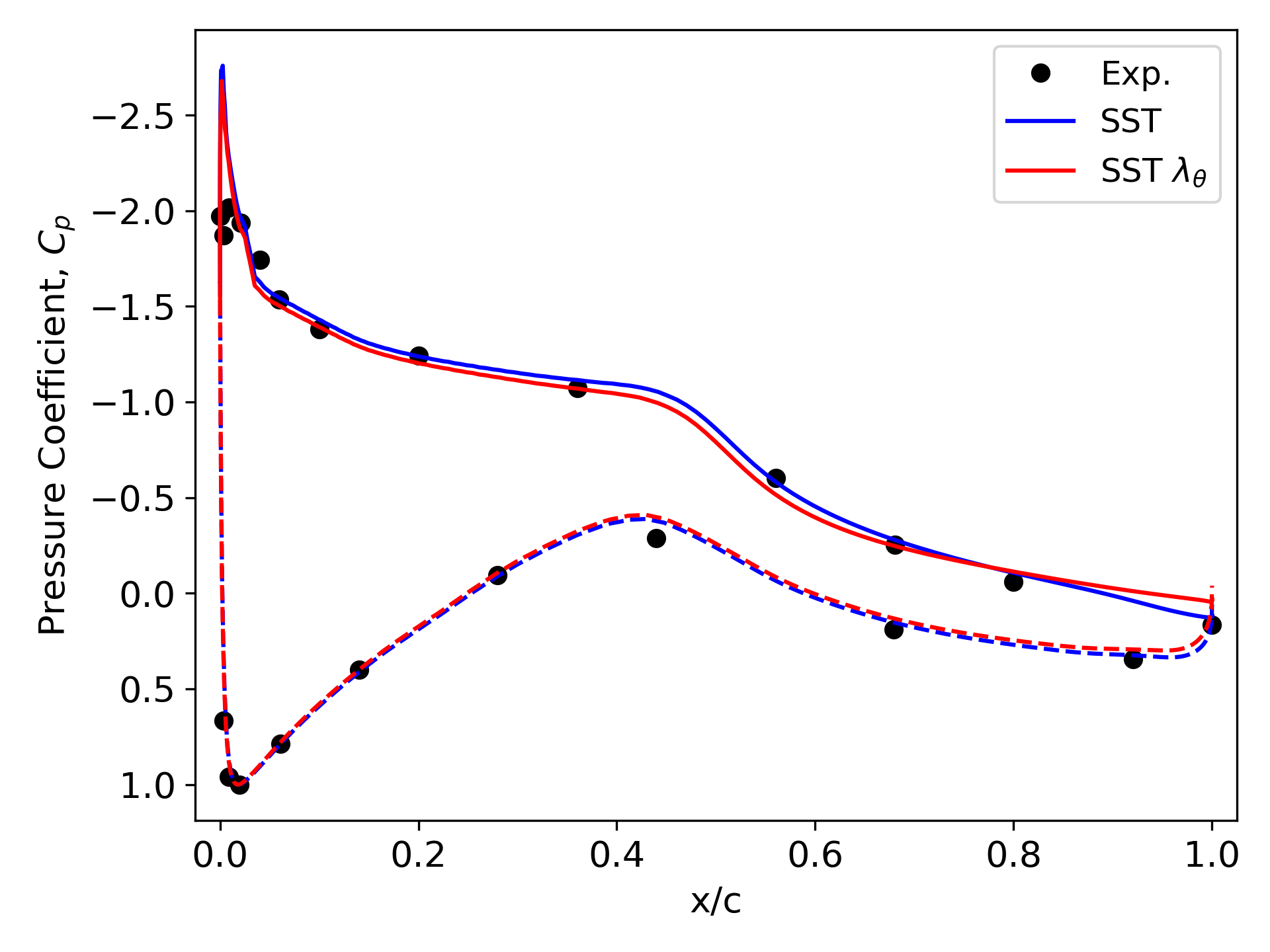}}
	\subcaptionbox{$r/R=0.80$}{\includegraphics[width=0.49\textwidth]{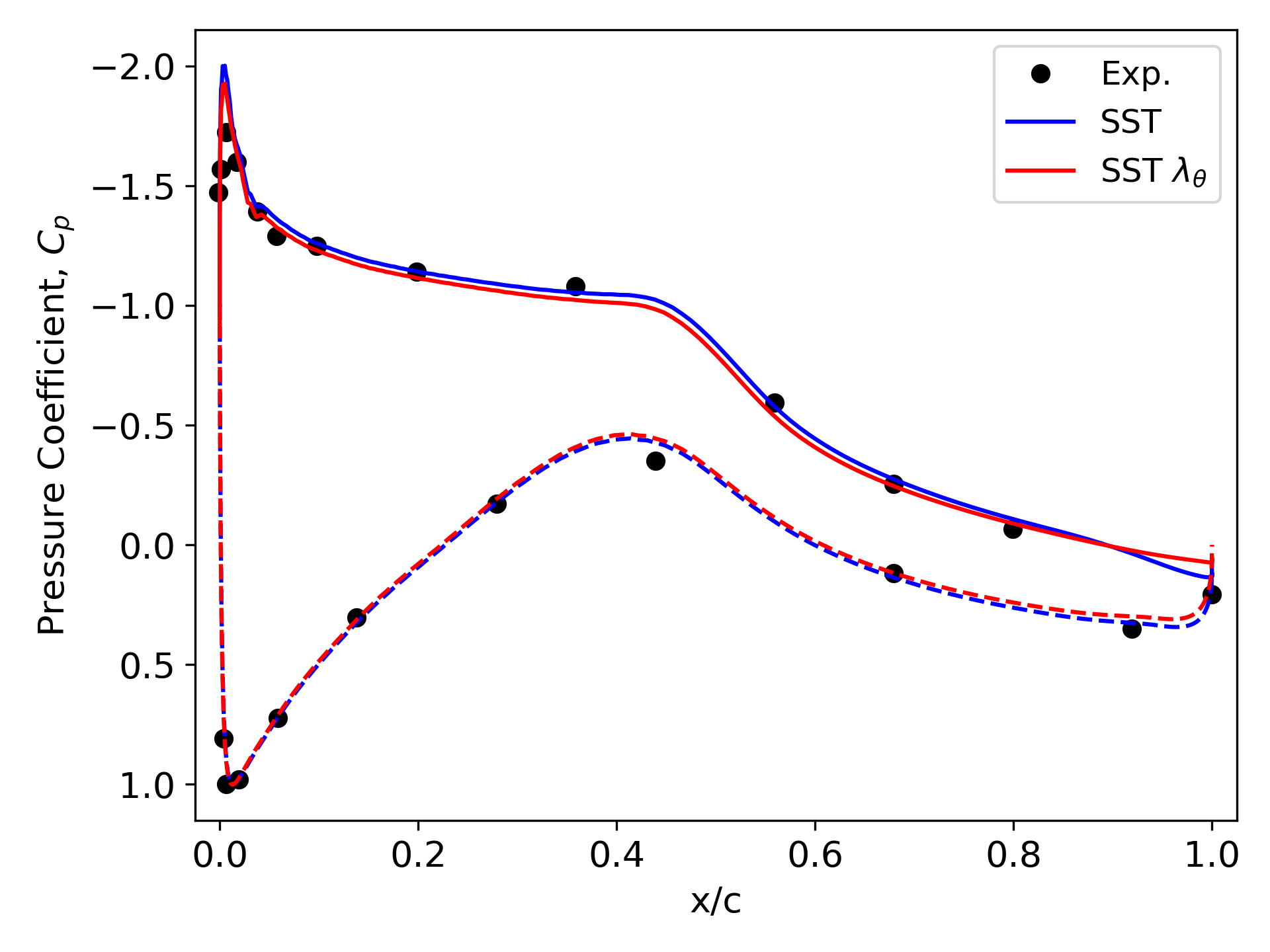}}
	\subcaptionbox{$r/R=0.95$}{\includegraphics[width=0.49\textwidth]{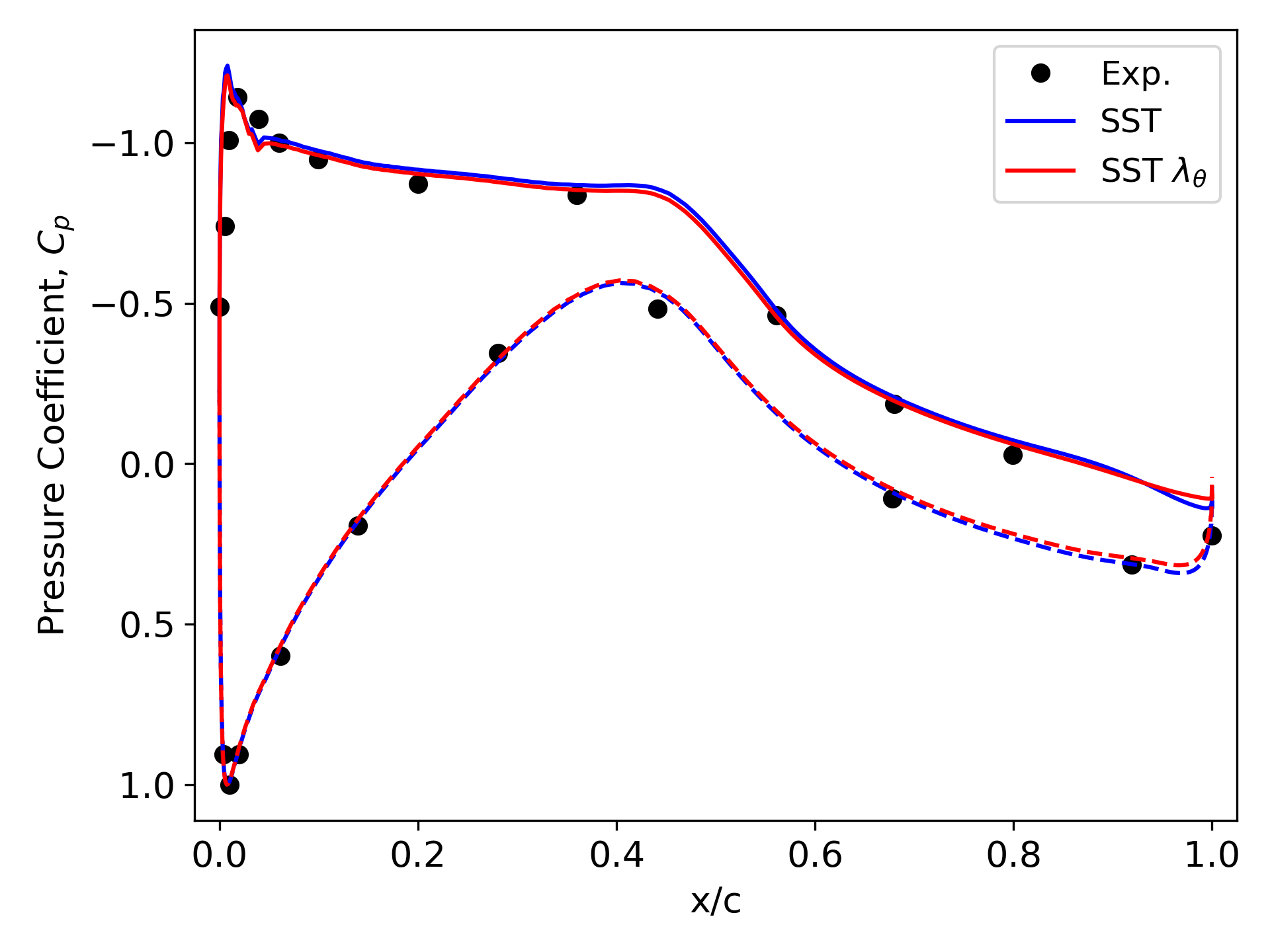}}
	\caption{Comparison of the pressure coefficient at four radial locations on the NREL Phase VI wind turbine blade. Solid/Dashed line: Suction/Pressure side, respectively.}
	\label{fig:phase_vi_cp}
\end{figure}

\begin{figure}[h]
	\centering
	\floatsetup[figure]{style=plaintop} 
    \includegraphics[width=0.6\textwidth]{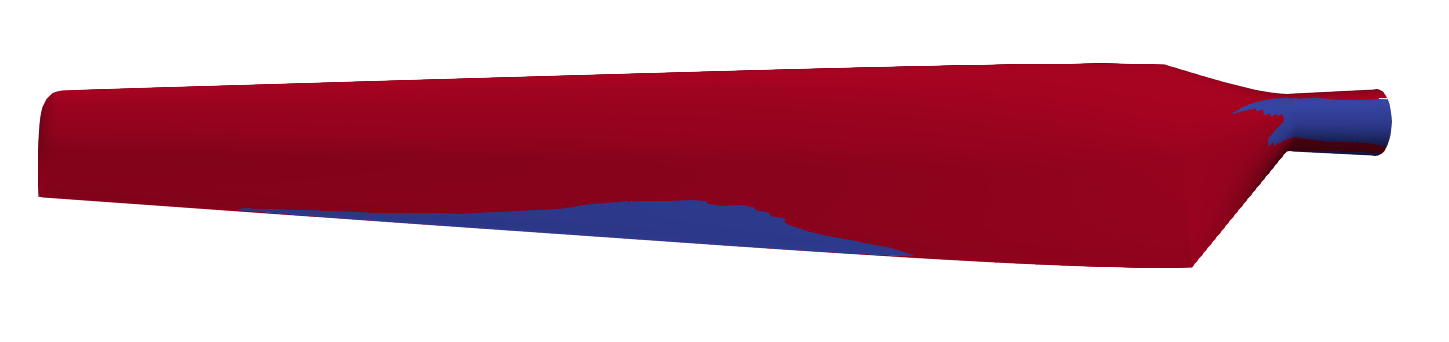}
	\caption{Attached (red) and separated (blue) flow regions on the suction side of the NREL Phase VI wind turbine blade. Obtained from the SST–$\lambda_{\theta}$ simulation.}
	\label{fig:phase_vi_blade_flow}
\end{figure}

Figure~\ref{fig:phase_vi_cp} presents comparisons of the surface pressure distributions between $r/R = 0.47$ and $r/R = 0.95$ on the blade surface. The comparisons are made among the experimental data~\citep{Hand2001}, the SST results, and the SST $\lambda_{\theta}$ results. The differences between the results with and without the separation correction are primarily observed on the suction side (corresponding to the upper surface of two-dimensional airfoils). Overall, the suction-side pressure coefficients show a lower magnitude over most of the surface when the correction is applied, followed by a slightly higher suction magnitude very near the trailing edge. This behavior leads to the reduction in the total thrust compared with the results without the separation correction, as presented in Figure~\ref{fig:phase_vi}. The observed trends are also consistent with the reduction of the lift and change in the pressure distributions observed in the S809 airfoil results shown in Figures~\ref{fig:s809} and~\ref{fig:s809_cpcf}. Figure~\ref{fig:phase_vi_blade_flow} shows the detailed flow state on the suction side of the blade surface obtained from the SST $\lambda_{\theta}$ simulation. In the figure, red and blue indicate the attached and separated flow regions, respectively. While the baseline SST simulation (not shown) exhibits no separation on the suction side, the separation-corrected results show separated flow regions between $r/R = 0.42$ and $r/R = 0.87$ near the trailing edge, which resulted in the differences in suction-side pressure observed in Figure \ref{fig:phase_vi_cp}.

Overall, the NREL Phase VI rotor case provides preliminary evidence that the model generalizes from two dimensional airfoils to three-dimensional rotors with sweep, twist, and rotation effects.

\section{Conclusion} \label{sec:conclusion}
We proposed a pressure-gradient-based sensor that is used to improve the performance of the SST turbulence model in flows with strong adverse pressure gradients (APGs). In addition, the model does not degrade the satisfactory performance of the baseline model in attached flows with neutral or favorable pressure gradients. The model performs accurately in both transitional and fully turbulent flows. The model is fully local; thus, it can be readily implemented, and it does not depend on any arbitrary length scales or velocity scales, unlike the antecedent model \citep{griffin2025improved}. The proposed model is applied to five airfoils between 12\% and 24\% thickness, including cases that are representative of wind turbines and aircraft with and without forced transition. For all cases, the proposed model preserves or improves the prediction of lift and drag, particularly at angles of attack near and beyond the onset of stall (large-scale separation) compared to the predictions of the baseline SST model. The models are also applied to the NREL Phase VI rotor, and the proposed model also outperforms the baseline model in this three-dimensional case involving blade sweep, twist, and rotation. Cases tested include turbulence intensities between 0.04 and 0.15\%, typical of applications to external aerodynamics and wind energy. The underlying transition model was designed and calibrated for turbulence intensities up to 6\%, so the outlook for deploying the proposed model in high-turbulence environments is favorable, though it remains to be tested. The model has been tested at chord-based Reynolds numbers between 1.5 and 4 million. The underlying transition model is known to exhibit degraded accuracy above $Re_c \approx$ $6\textendash9$ million. Thus, in high-Reynolds-number applications without a fixed transition location, the proposed sensor may need to be deployed with a more suitable transition model. This study was limited to RANS simulations with large time steps, which are suitable for predicting steady flow phenomena associated with the onset of stall. Accurate predictions of large-scale unsteady separations including deep stall will be the study of future work and the present model is suitable for integration with time-resolving turbulence modeling approaches for this purpose. This preliminary investigation suggests the model generalizes across applications without additional tuning and merits further study. 

\appendix
\section{Definitions of the production and destruction/relaminarization source terms} \label{app:source_terms}
$P_\gamma$ is the production source term of intermittency as defined in Menter et al. \citep{Menter2015} as
\begin{equation}
    P_\gamma = F_{length} \rho S \gamma (1-\gamma) F_{onset},
\end{equation}
where $F_{length} = 100$ and
\begin{equation}
    F_{onset} = \max(F_{onset2}-F_{onset3},0),
\end{equation}
\begin{equation}
F_{onset3} = \max \left( 1 - \left(\frac{R_T}{3.5}\right)^3, 0 \right),
\end{equation}
with $R_T=\rho k / (\mu \omega)$, $F_{onset2} = \min(F_{onset1}, 2.0)$
\begin{equation}
    F_{onset1} = \frac{Re_{\mathrm{v}}}{2.2 Re_{\theta c}},
\end{equation}
$Re_{\mathrm{v}} = \rho d_w^2 S / \mu$, $Re_{\theta c} = C_{TU1} + C_{TU2} e^{-C_{TU3} Tu_L F_{PG}}$, $C_{TU1}=100.0$, $C_{TU2}=1000.0$, $C_{TU3}=1.0$,
\begin{equation}
    Tu_L = \min(100\frac{\sqrt{2k/3}}{\omega d_w}, 100),
\end{equation}
\begin{equation}
F_{PG}^\prime = \begin{cases}
\min(1 + C_{PG1}\lambda_\theta, C_{PG1}^\mathrm{lim}) & \lambda_\theta \ge 0\\
\min(1 + C_{PG2}\lambda_\theta, C_{PG2}^\mathrm{lim}) & \lambda_\theta < 0,
\end{cases}
\end{equation}
$F_{PG} = \max(F_{PG}^\prime,0)$, $C_{PG1} = 14.68$, $C_{PG2} = -7.34$, $C_{PG1}^\mathrm{lim} = 1.5$, \text{and} $C_{PG2}^\mathrm{lim} = 3.0$.

$E_\gamma$ is the destruction/relaminarization source term of intermittency defined in Menter et al. \citep{Menter2015} as
\begin{equation}
    E_\gamma = c_{a2} \rho \Omega \gamma F_{turb} (c_{e2}\gamma - 1),
\end{equation}
with $F_{turb} = e^{-(R_T/2)^4}$, $c_{a2} = 0.06$, \text{and} $c_{e2} = 50$.

\section{Grid and time-step convergence} \label{app:grid_conv}
Refinement studies of the sensitivity to spatial grid size and time step are performed to assess the convergence of the solution with respect to these parameters. As a baseline, we consider the the fully turbulent SST model for the DU00-w-212 airfoil as was presented in Figure \ref{fig:du00w212_R03M}. We consider individually two types of refinement: spatially refining the wall-normal grid size in the wall-normal direction by a factor of two and reducing the time step size by a factor of two. The results for grid size refinement are shown in Figure \ref{fig:du00w212_conv}(a,b) and the results for time-step refinement are shown in Figure \ref{fig:du00w212_conv}(c,d). In both cases, the red symbols indicate the results with the baseline parameters, and the blue symbols indicate the results with the refined parameters. Both the lift and drag coefficients exhibit convergence under both grid-size and time-step refinement. Since the time step and grid parameters are the same for all airfoils considered in this study, this convergence study is assumed to be representative for the other airfoils considered in this work. We direct the reader to \cite{lee2025assessment} for more comprehensive convergence studies for the S809 and NLF(1)-0416 airfoils using our solver and the grids deployed in this study as well as five additional resolutions. That study also includes cross-verification of our solver with NASA's OVERFLOW and FUN3D solutions with identical turbulence and transition models, mesh resolution, and $\mu_t/\mu$ ratio.
\begin{figure}[h]
	\centering
	\floatsetup[figure]{style=plaintop} 
	\subcaptionbox{}{\includegraphics[width=0.49\textwidth]{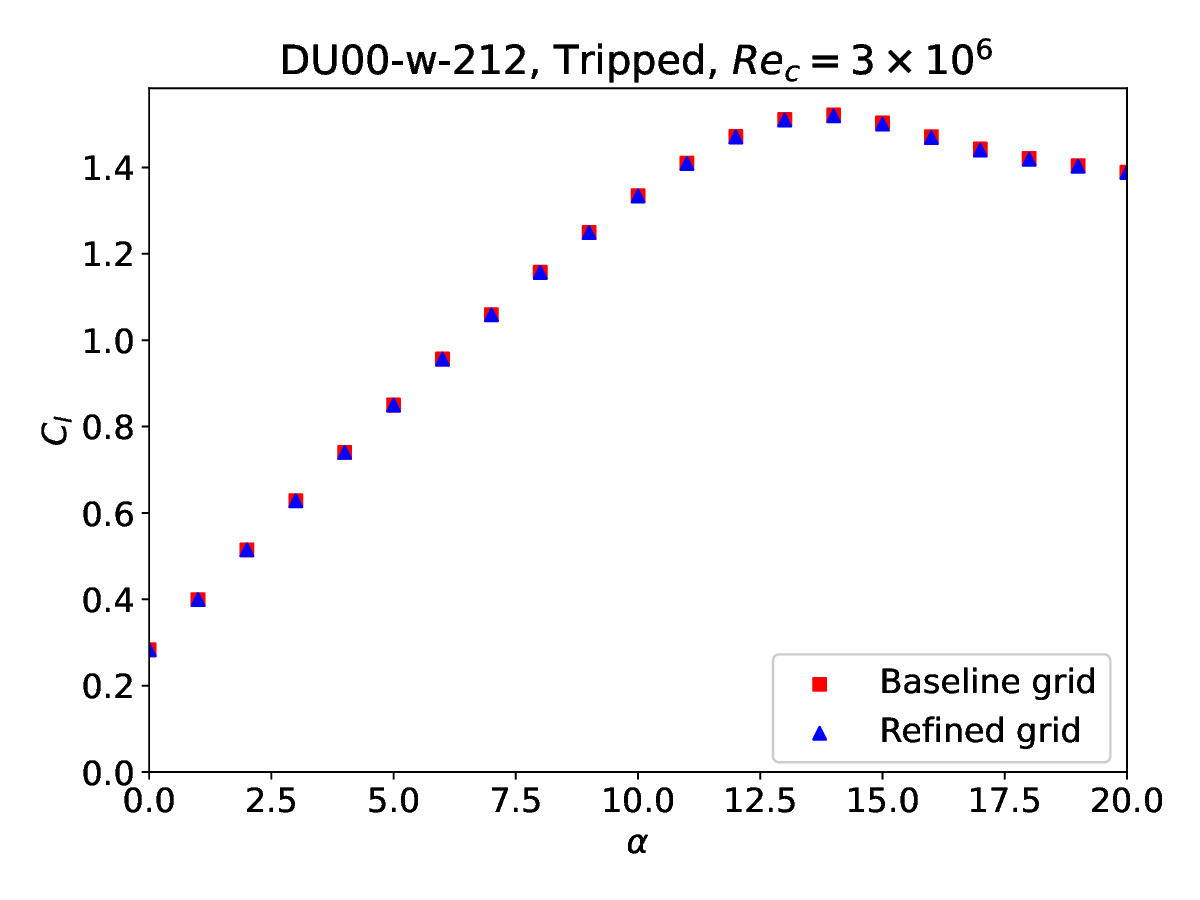}}
	\subcaptionbox{}{\includegraphics[width=0.49\textwidth]{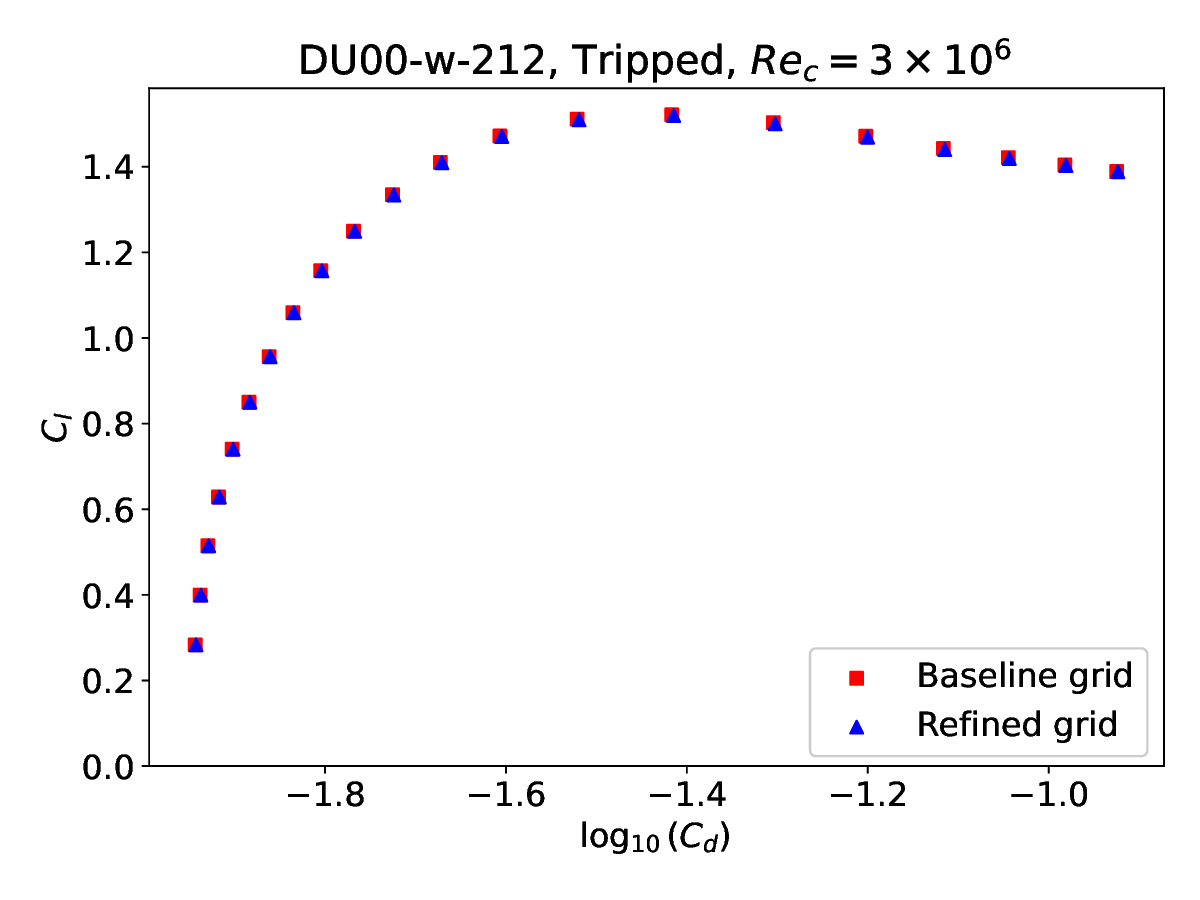}}
	\subcaptionbox{}{\includegraphics[width=0.49\textwidth]{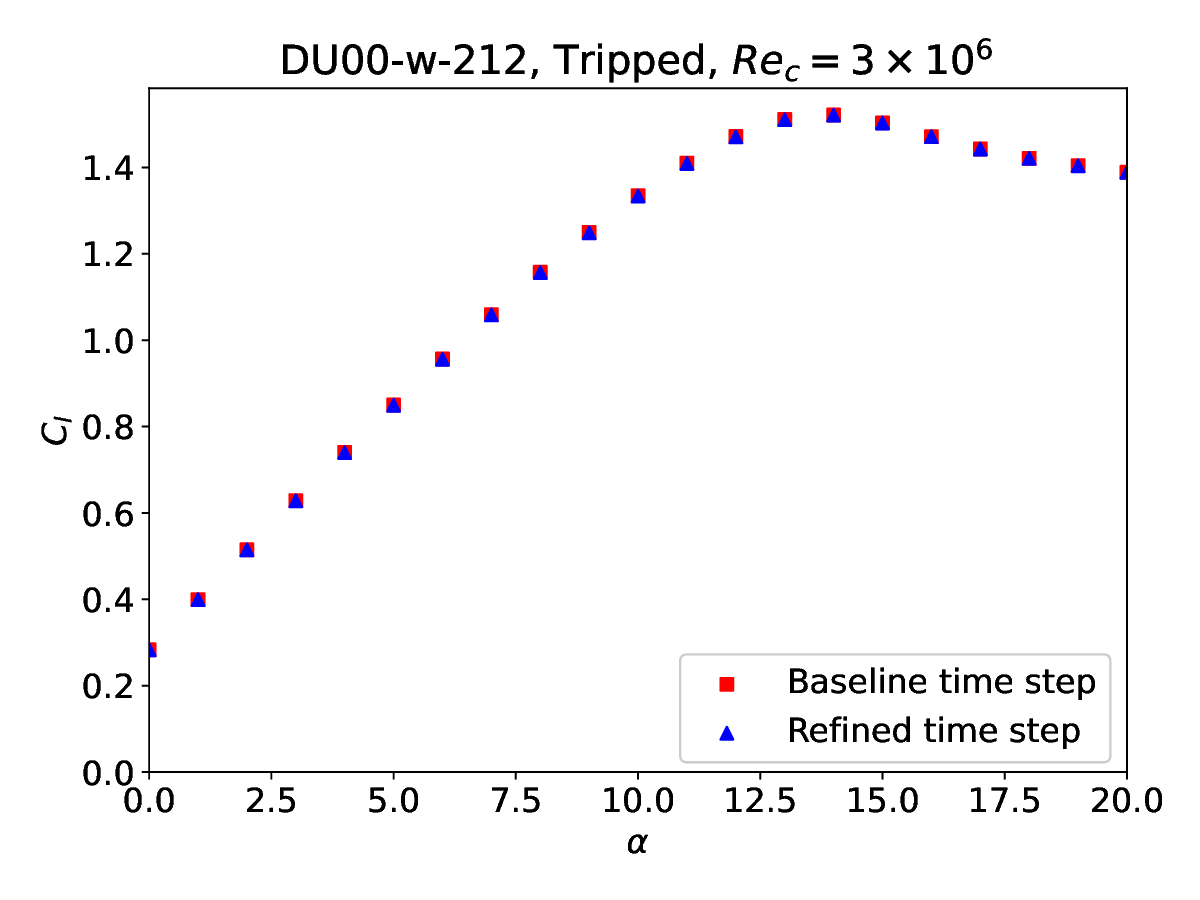}}
	\subcaptionbox{}{\includegraphics[width=0.49\textwidth]{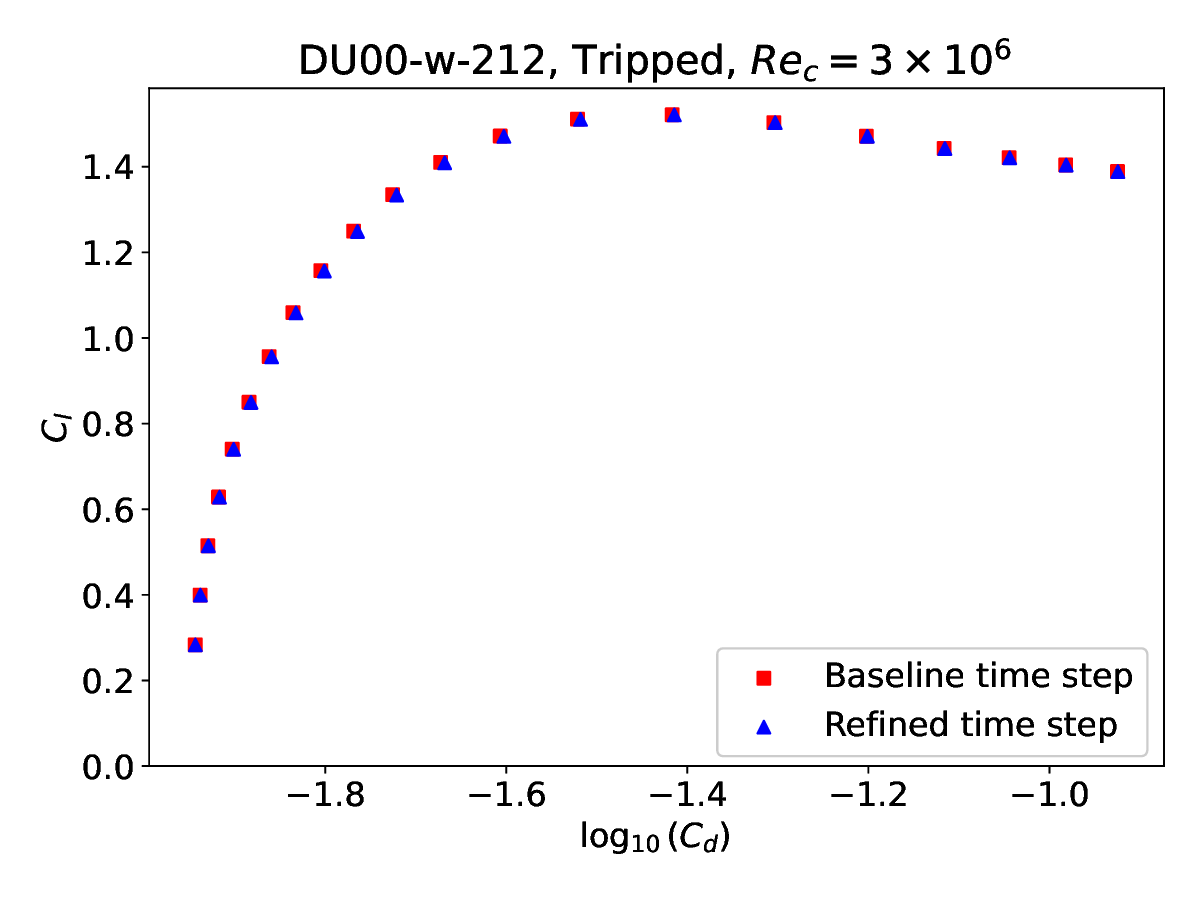}}
	\caption{Coefficient of lift versus (a,c) angle of attack and (b,d) coefficient of drag for the DU00-w-212 airfoil as predicted by the fully turbulent (tripped) SST model. Grid convergence (a,b) and time step convergence (c,d) are assessed. The results for the baseline grid and time steps are shown in red, and results with either the grid or time step refined are shown in blue.}
	\label{fig:du00w212_conv}
\end{figure}

\section{Temporal convergence of forces} \label{app:force_conv}
The solver is temporally advanced for 10,000 time steps with a time-step size of 2 steps per chord-pass time. The temporal evolution of the coefficients of lift and drag are shown in Figure \ref{fig:du00w212_conv} for the flow over the DU00-w-212 at various angles of attack. For all cases, a steady-state solution is established indicating that adequate time advancement has occurred. Throughout the manuscript, time-averaged results are presented by averaging the samples from time steps 6,000 through 10,000 since the lift and drag coefficients are essentially constant over this range. A steady solution for separated flows is expected for RANS calculations since schemes with significant numerical dissipation are conventional. The present numerics use upwinding thus are no exception to convention.

\begin{figure}[h]
	\centering
	\floatsetup[figure]{style=plaintop} 
	\subcaptionbox{}{\includegraphics[width=0.49\textwidth]{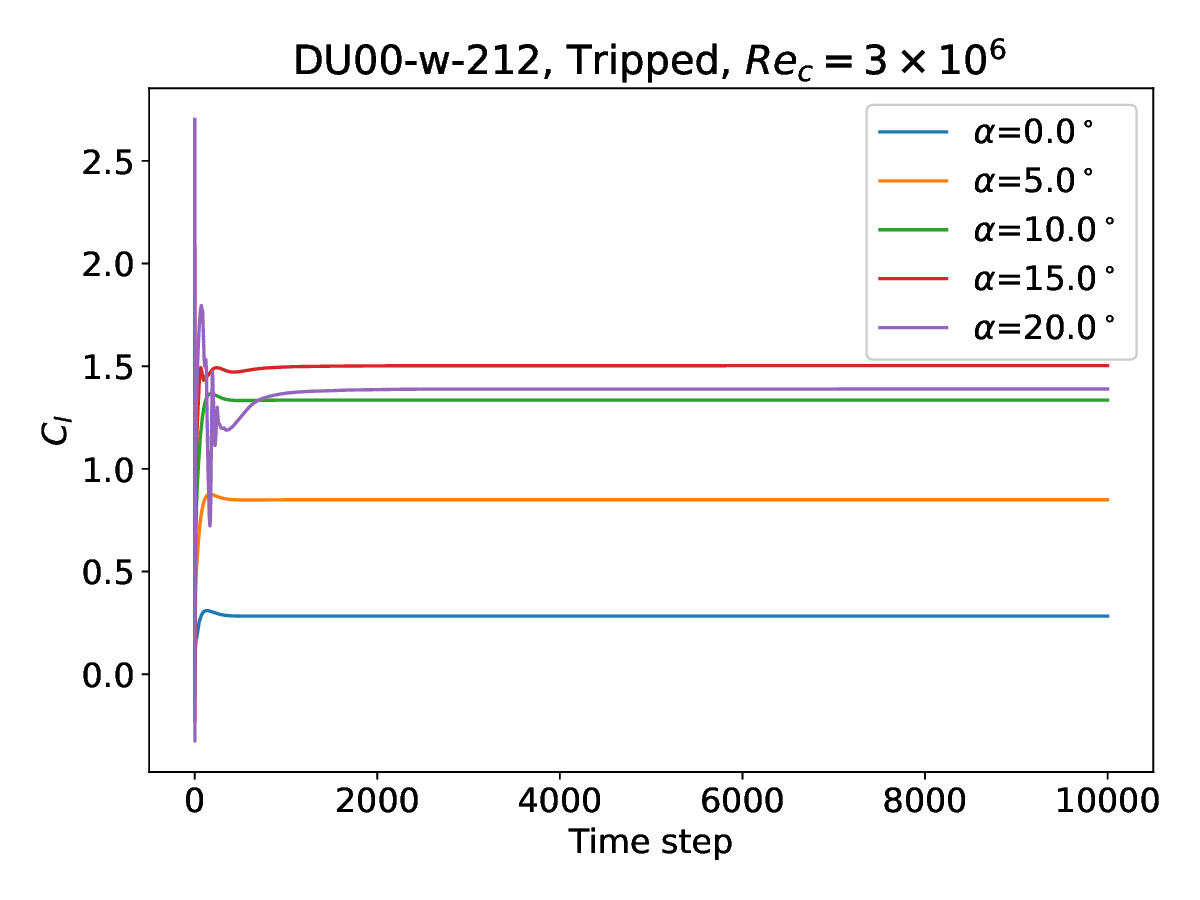}}
    \subcaptionbox{}{\includegraphics[width=0.49\textwidth]{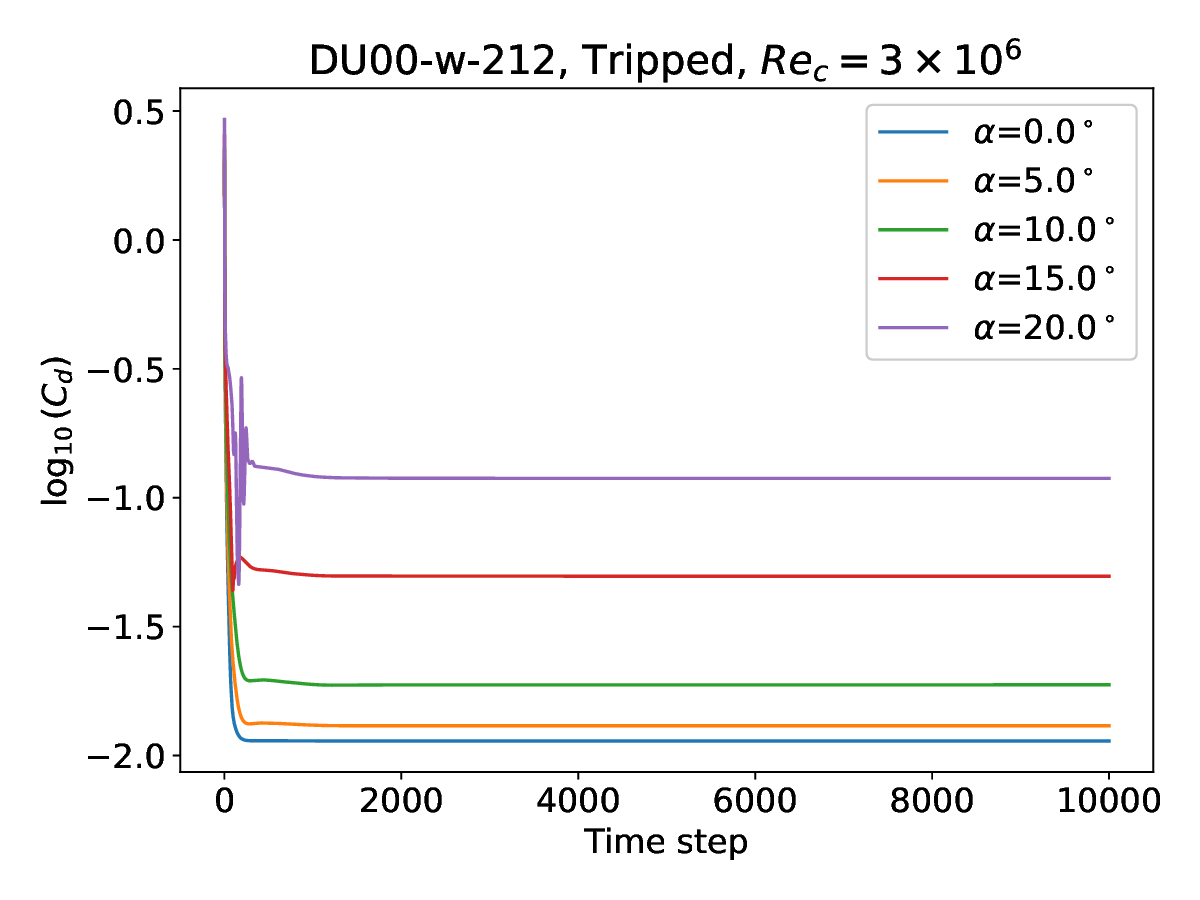}}
	\caption{Convergence in time of lift (a) and drag (b) coefficients. Five angles of attack are included for the DU00-w-212 airfoil using the fully turbulent (tripped) SST model.}
	\label{fig:du00w212_conv}
\end{figure}

\section*{Data Availability Statement}
The data that support the findings of this study are available from the corresponding author upon reasonable request.

\section*{Funding Sources}
Funding provided by the U.S. Department of Energy Office of Energy Efficiency and Renewable Energy, Wind Energy Technologies Office. Benjamin Savino was supported by the U.S. Department of Energy Office of Science Graduate Student Research Program.

\section*{Acknowledgments}
We would like to acknowledge insightful discussions at the Center for Turbulence Research Summer Program 2024, especially with Matteo Dellacasagrande. 
This work was authored in part by the National Laboratory of the Rockies for the U.S. Department of Energy (DOE) under Contract No. DE-AC36-08GO28308. The views expressed in the article do not necessarily represent the views of the DOE or the U.S. Government. The U.S. Government retains and the publisher, by accepting the article for publication, acknowledges that the U.S. Government retains a nonexclusive, paid-up, irrevocable, worldwide license to publish or reproduce the published form of this work, or allow others to do so, for U.S. Government purposes. The research was performed using computational resources sponsored by the U.S. Department of Energy’s Office of Critical Minerals and Energy Innovation and located at the National Laboratory of the Rockies.

\section*{Declaration of Competing Interests}
The authors declare no competing interests.

\bibliographystyle{unsrt}

\bibliography{references}

\end{document}